\documentclass[11pt]{article}

\usepackage[utf8]{inputenc}
\usepackage{graphicx}
\usepackage[colorlinks,allcolors=cyan]{hyperref}
\usepackage{amsmath,amssymb,amsfonts,mathtools}

\usepackage{lipsum}
\usepackage{wrapfig}
\usepackage{graphicx,overpic}
\usepackage{siunitx}
\usepackage{textcomp}
\usepackage{titlesec}

\usepackage{subcaption}
\usepackage{graphicx}
\graphicspath{{img/}} % includegraphics path
\newcommand{\includegraphicsw}[2][1.]{\includegraphics[width=#1\linewidth]{#2}}
\usepackage{multirow}
\usepackage{hhline}

\usepackage{bm}
\newcommand{\vect}[1]{\boldsymbol{\mathbf{#1}}}
\newcommand{\bu}{\mathbf u}

\newcommand*\diff{\mathop{}\!\mathrm{d}}

\newcommand{\gradG}{\nabla_{\Gamma}}
\newcommand{\laplG}{\Delta_{\Gamma}}

\newcommand{\divG}{{\mathop{\,\rm div}}_{\Gamma}}

\begin{document}

\title{Lipid domain coarsening and fluidity in multicomponent lipid vesicles: A continuum based model and its experimental validation}
%Validation of a continuum based model of coarsening and fluidity in multicomponent lipid vesicles

\author{Y.~Wang$^{1,\dag}$, Y.~Palzhanov$^{2,\dag}$, A.~Quaini$^{2}$, M.~Olshanskii$^{2,*}$, S.~Majd$^{1,*}$}

\maketitle

\begin{center}
\noindent $^{1}$Department of Biomedical Engineering, University of Houston, 3551 Cullen Blvd, Houston TX 77204\\
\texttt{ywang147@uh.edu; smajd9@central.uh.edu}

\vskip .3cm
\noindent $^{2}$Department of Mathematics, University of Houston, 3551 Cullen Blvd, Houston TX 77204\\
\texttt{ypalzhanov@uh.edu; aquaini@uh.edu; maolshanskiy@uh.edu}
\end{center}

\vskip .3cm
\noindent $^{\dag}$ Equal contribution \\
$^{*}$ Corresponding authors

\vskip .3cm
\noindent{\bf Abstract}
Liposomes that achieve a heterogeneous and spatially organized surface through phase separation
have been recognized to be a promising platform for delivery purposes. However, 
their design and optimization through experimentation can be expensive and time-consuming.
To assist with the design and reduce the associated cost, we propose a computational platform 
for modeling membrane coarsening dynamics based on the  principles of continuum mechanics and thermodynamics. This model couples phase separation to lateral flow and accounts for different membrane fluidity within the different phases, which is known to affect the coarsening dynamics on lipid
membranes. The simulation results  are in agreement with the experimental data in terms of liquid ordered domains area fraction,
total domains perimeter over time and total number of domains over time for two different membrane compositions
(DOPC:DPPC with a 1:1 molar ratio with 15\% Chol and DOPC:DPPC with a 1:2 molar ratio with 25\% Chol) that yield 
opposite and nearly inverse phase behavior. This quantitative validation shows that the developed platform
can be a valuable tool in complementing experimental practice. 


\section{Introduction}

Biological membranes are heterogenous and this characteristic is critical for their functionality. 
The lipid bilayer in these membranes hosts a variety of lipid species that may be organized into one 
of the two phases: liquid disordered and liquid ordered \cite{Balint2017}. 
The tight packing of saturated lipids and cholesterol in the liquid ordered phase, in contrast to the loosely 
packed unsaturated lipids present in the liquid disordered phase leads, under certain conditions, to the lipid phase separation in membranes \cite{Brown1}. The liquid ordered domains  - also known as lipid rafts - that are surrounded 
with liquid disordered phase in biological membranes, have been recognized as a key platform for cell 
signaling and membrane trafficking among other cellular processes \cite{Brown1,brown2000,Lingwood46,Bandekar2012,Trementozzi2019}.
Thus, these domains have received growing interest in the past few decades and have been the focus of numerous 
experimental and theoretical studies \cite{Levental2020,BENNETT20131765}.

More recently, domain formation on membranes has also been utilized to create novel membrane-based 
materials with heterogenous surfaces. When explored for drug delivery applications, these heterogenous 
membrane materials showed clear advantage over their homogenous membrane counterparts \cite{Bandekar2013,Sempkowski2016}. 
With the increasing number of available lipid-conjugated molecules (e.g. peptides, polymers, etc), 
lipid membranes with heterogenous and spatially-organized surfaces can open new avenues for the design of novel materials.  
However, efficient design of such heterogenous membrane-based materials requires computer-aided modeling that can predict 
the lipid domain formation and dynamics on a given membrane composition in a reliable and quantitative manner. 
As a step towards addressing this need, we recently developed a computationally efficient method based on
the surface Cahn--Hilliard phase-field model to predict the phase behavior and domain formation on ternary membrane 
compositions and we validated this continuum-based model using experiments \cite{zhiliakov2021experimental}. In this study, 
we aim to further advance this computational platform to offer an enhanced and more accurate prediction of membrane phase separation.

Our current understanding of membrane phase separation is mainly based on the experimental studies 
performed on model membranes with well-defined lipid compositions. Amongst these model membranes, 
giant unilamellar vesicles (GUVs) have provided a particularly suitable platform for studying membrane 
phase behavior as their free-standing lipid bilayer closely mimics the natural membranes and their large size 
(micron-scale) makes them resolvable under optical microscopy \cite{Wesolowska2009}. 
The combinatorial use of GUVs and advanced fluorescence-based microscopy techniques have, 
for instance, shaped our knowledge on membrane domains' thermodynamic equilibria \cite{Veatch17650,FIDORRA20092142}
and their coarsening dynamics \cite{Stanich2013}. These studies have also provided us with an 
insight into the distinct characteristics (e.g. morphology and fluidity) of liquid ordered and liquid disordered phases
\cite{Kahya2003,B901587F,SEZGIN20121777}. 

It has been demonstrated that membrane fluidity within the liquid ordered 
domains can be substantially lower than that in the liquid disordered phase \cite{SEZGIN20121777}.  
Such a difference can affect the coarsening dynamics of rafts on membranes \cite{Stanich2013}. 
This interesting aspect of lipid domains was not considered in our previous study \cite{zhiliakov2021experimental}
as fluidity is not accounted for in the Cahn--Hilliard model. 
In order to enable more accurate predictions on membrane phase behavior with consideration of membrane 
viscosity differences between liquid ordered and disordered phases, herein we apply a more complex model, 
the Navier--Stokes--Cahn--Hilliard model, along with a set of innovative numerical methods, to present an
advanced computational platform for modeling membrane phase separation. To validate this model, we compare 
its numerical results to our experimental results on GUVs with ternary membrane compositions. Specifically, 
we apply fluorescence microscopy to examine the phase separation on GUVs with two distinct compositions 
(with opposite and nearly inverse phase behavior) and monitor the number of lipid domains, area fraction 
and perimeter over time and compare these results to those from our computational model. 

\section{Materials and methods}

\subsection{Experimental approach}

\subsubsection{Materials}

Lipids 1,2-dioleoyl-sn-glycero-3-phosphocholine (DOPC), 1,2-dipalmitoyl-sn-glycero-3-phosphocholine (DPPC), 
1,2-dipalmitoyl-sn-glycero-3-phosphoethanolamine-N- (lissamine rhodamine B sulfonyl) (Rho-PE) 
were purchased from Avanti Polar Lipids (Alabaster, AL). We purchased the sucrose from VWR (West Chester, PA). 
Cholesterol was from Sigma Aldrich (Saint Louis, MO) and chloroform from Omnipure (Caldwell, Idaho). 
Naphtho[2,3-a]pyrene (NAP) was from Thermo Fisher Scientific (Waltham, MA). All lipid stock solutions 
were prepared in chloroform. Indium tin oxide (ITO) coated glasses and microscope glass slides were from 
Thermo Fisher Scientific (Waltham, MA) and coverslips were from Corning Inc. (Corning, NY). ITO plates 
were cleaned using chloroform, ethanol and DI water prior to use. Microscope slides and coverslips were cleaned 
with ethanol and DI water before usage. 

\subsubsection{Preparation of Giant Unilamellar Vesicles (GUVs)}

GUVs were prepared using a modified version of electroformation technique described in our previous studies \cite{majd2005hydrogel,park2018reconstitution,zhiliakov2021experimental}. In brief, an aqueous dispersion of 
small vesicles was used to produce dried lipid films on two ITO plates that later sandwiched a thin PDMS frame, 
creating an electroformation chamber. A sucrose solution (235 mM) was injected into the chamber to rehydrate the lipid films. 
Next, an AC electric field was applied while the chamber was placed in an oven ($\sim$60\textdegree{}C) to 
exceed the melting temperature of the used lipids. With a 50 Hz frequency, the electric field was increased to 2 
Vpp at rate of 100 mVpp/min and held for $\sim$3 hours using a function waveform generator (4055, BK Precision, Yorba Linda, CA) 
upon formation, GUVs were detached by decreasing the frequency to 1 Hz for $\sim$30 min. 

The above-mentioned small vesicles were prepared using dehydration-rehydration followed by tip-sonication as described 
in our prior studies \cite{kang2013simple,zhiliakov2021experimental}. A mixture of DOPC, DPPC, and Chol plus 0.3 mol\% 
Rho-PE and 0.5\% NAP was prepared in chloroform and used to produce a thin lipid film in a pearl-shaped flask using a 
rotary evaporator (Hei-Vap, Heidolph, Germany). The lipid film was then rehydrated using DI water (at ($\sim$60\textdegree{}C) 
with a 2.5 mg/ml lipid concentration. The resultant milky suspension was tip-sonicated to produce a clear suspension of small vesicles. 

Lipid compositions applied in this study were (i) DOPC: DPPC with a 1:1 molar ratio and 15\% Chol, referred to as 1:1:15\% 
composition, and (ii) DOPC: DPPC with a 1:2 molar ratio and 25\% Chol, referred to as 1:2:25\% composition. We included 
Rho-PE and NAP in both compositions to enable fluorescence microscopy.

\subsubsection{GUV Imaging and Analysis}

GUV imaging and characterization was performed as described before \cite{zhiliakov2021experimental}. 
Briefly, GUVs were harvested from the electroformation chamber and placed on a clean microscope glass slide for imaging. 
Images were acquired using a Zeiss LSM 800 confocal laser scanning microscope (Zeiss, Germany). Prior to imaging, 
the sample was heated on a hot plate to $\sim$60\textdegree{}C for 5 min and then placed on the microscope stage 
where it gradually cooled down to the room temperature. The image collection time was recorded with time zero 
considered as when the sample was removed from the hot plate. Epi-fluorescence microscopy was used for the 
initial assessment of GUVs and their lipid domains while confocal microscopy was used to further assess the domains 
on GUVs and quantify their size. Epi-fluorescence images were collected using a 40X objective with numerical aperture 
(NA) of 0.95, with 545/488 excitation wavelength and 572/509 emission wavelength. 
Confocal images were collected using a 63X oil objective with NA of 1.40 using 488 nm and 561 nm wavelength laser. 
Confocal image slices were collected with 0.4-\SI{0.9}{\micro\metre} Z-steps, depending on the size of the examined 
GUV, to minimize the time required for imaging the entire vesicle without significant movement of vesicle. 
Confocal images were analyzed using ZEN software (ZEN 2.6 lite, Zeiss, Germany). 

Considering that Rho-PE and NAP have shown preferential partitioning to the liquid disordered phase and 
liquid ordered phase \cite{PMID:20642452,KLYMCHENKO201497}, respectively, we assumed that green patches 
on the examined GUVs represented the liquid ordered phase while the red regions represented the liquid disordered phase. 

For image-based characterization of GUVs, we assumed that the GUVs were perfect spheres and that the lipid 
domains on GUVs were in the form of spherical caps. For each GUV, all collected confocal image slices of each 
GUV were analyzed to find the vesicle diameter and surface area, as well as the surface area and perimeter 
of the lipid domains as detailed in \cite{zhiliakov2021experimental}. We performed this analysis on 20 GUVs with 
1:2:25\% composition and 18 GUVs with 1:1:15\% composition from at least 3 independent experiments per composition. 

\subsection{Computational approach}

\subsubsection{Mathematical model}\label{sec:math_model}

In order to model phase separation, viscous and fluidic phenomena occurring in the GUV membrane, 
we consider a suitable modification \cite{Palzhanov2021} of a well-known thermodynamically consistent Navier--Stokes--Cahn--Hilliard (NSCH) model \cite{Abels2012}. 
Let $\Gamma$ be a sphere representing a liposome with a
\SI{10}{\micro\metre} diameter and let $c_i$ be a fraction of elementary surface area occupied by phase $i$, with $i = 1, 2$.
We choose $c = c_1$, $c\in [0,1]$, as the representative surface fraction, e.g.~the fraction of the ordered phase. 
Let $\rho_1$ and $\rho_2$ be the densities and $\eta_1$ and $\eta_2$ the dynamic viscosities of the two phases.
Then, the density and viscosity of the mixture can be written as $\rho= \rho(c) = \rho_1 c+ \rho_2 (1-c)$
{and} $\eta=\eta(c)=\eta_1 c+\eta_2(1-c)$. Finally, let $\bu$ be the averaged
tangential velocity in the mixture, $p$ the thermodynamic interfacial pressure, and $\mu$ the chemical potential.

The NSCH system that governs the evolution of $c$, $\bu$, $p$, and $\mu$ in time $t$ and space $\vect x \in\Gamma\subset\mathbb R^3$
is given by:
\begin{align}
& \underbrace{\rho(\partial_t \bu + (\gradG \bu)\bu)}_{\text{inertia}} - \underbrace{\operatorname{\mathbf{div}}_{\Gamma}(2\eta E_s(\bu))+ \gradG p}_{\text{lateral stresses}} =  \underbrace{-\sigma_\gamma \epsilon^2 \divG \left(  \gradG c \otimes\gradG c \right)}_{\text{line tension}} + \underbrace{{M \theta(\gradG(\theta\bu)\,)\gradG \mu}}_{\text{chemical momentum flux}} \label{grache-1m} \\
& \underbrace{\divG \bu  =0}_{\text{membrane inextensibility}}&  \label{gracke-2}\\
&\underbrace{\partial_t c +\divG(c\bu)}_{\text{transport of phases}}-  \underbrace{\divG \left(M \gradG \mu \right)}_{\scriptsize\begin{array}{c}\text{phase masses exchange}\\ \text{Fick's law}\end{array}}  = 0,\qquad
\mu = \underbrace{ f_0'(c) - \epsilon^2 \laplG c}_{\text{mixture free energy variation}}  \label{gracke-4}
\end{align}
on $\Gamma$ for $t \in (0,t^{\rm final}]$. In eq.~\eqref{grache-1m}-\eqref{gracke-4}, $\nabla_\Gamma$ stands for the tangential gradient, $\Delta_\Gamma$ for the Laplace--Beltrami operator,   
$E_s(\bu) = \frac12(\nabla_\Gamma \bu + (\nabla_\Gamma \bu)^T)$ is the Boussinesq--Scriven strain-rate tensor, and
$\divG$ is the surface divergence. 
Eq.~\eqref{gracke-4} provides the definition of the chemical potential, with $f_0(c) = \frac{1}{4}\,c^2\,(1 - c)^2$
being the double-well thermodynamic potential and parameter~$\epsilon > 0$ representing
the width of the (diffuse) interface between the phases. In addition, $\sigma_\gamma$ is line tension coefficient, 
$M$ is the mobility coefficient (see \cite{Landau_Lifshitz_1958}), and 
$\theta^2 = \frac{d\rho}{dc}$. Problem \eqref{grache-1m}-\eqref{gracke-4} 
models the total exchange of matter between phases (eq.~\eqref{gracke-4}) 
with surface flow described in terms of momentum conservation (eq.~\eqref{grache-1m}) 
and area preservation (eq.~\eqref{gracke-2}). Finally, problem \eqref{grache-1m}--\eqref{gracke-4} needs to be
supplemented with initial values of velocity $\bu_0$ and state $c_0$. Here, we take $c_0=c_0(\vect x)$ 
corresponding to a homogeneous mixture and $\bu_0 = \boldsymbol{0}$ (surface fluid at rest).

Several experimental works help with the settings of viscosity \cite{SAKUMA20201576}
and line tension \cite{Heftberger2015,C3SM51829A,KUZMIN20051120}. 
For density, we calculated the value for each phase using the estimated molecular weight and molecular surface area for 
the corresponding phase (see supplementary material for details). 
Table \ref{tab:physical_param} reports the values or range of values for such parameters for both compositions under consideration at two
temperatures related to our experiments.

\begin{table}[htb]
\begin{center}
 \begin{tabular}{ | l |  l |  l |  l |  l |  l | }
\hline
Composition (Temp) & $\rho_{l_o}$ & $\rho_{l_d}$ & $\eta_{l_o}$ & $\eta_{l_d}$ & $\sigma_\gamma$ \\
\hline
1:2:25\% (15\textdegree{}C) & 14.67 & 11.75 & $5-8 \cdot 10^{-8}$ & $0.2-0.4\cdot 10^{-8}$ & $1.2-1.8$ \\
\hline
1:2:25\% (17.5\textdegree{}C) & 14.35 & 11.72& $5-8 \cdot 10^{-8}$ & $0.2-0.4\cdot 10^{-8}$ & $1.2-1.8$ \\
\hline
1:1:15\% (22.5\textdegree{}C) & 14.01 & 11.72 & $0.43-5.7\cdot 10^{-8}$ & $0.2-0.4\cdot 10^{-8}$ & $1.2-1.6$ \\
\hline
1:1:15\% (25\textdegree{}C) & 13.69 & 11.84 & $0.43-5.7\cdot 10^{-8}$ & $0.2-0.4\cdot 10^{-8}$ & $1.2-1.6$ \\
\hline
\end{tabular}
\caption{Calculated value or range of values for the density of liquid ordered ($\rho_1=\rho_{l_o}$) and liquid disordered ($\rho_2=\rho_{l_d}$) phases in g/(mol$\cdot$\r{A}$^2$), viscosity of liquid ordered ($\eta_1=\eta_{l_o}$) and liquid disordered ($\eta_2=\eta_{l_d}$) phases in Pa$\cdot$s$\cdot$m, and line tension in pN for the two membrane compositions at different temperatures.
See supplementary material for details on the density calculations.
}\label{tab:physical_param}
\end{center}
\end{table}

%, $\lbrack \epsilon \rbrack = \unitLength$. For the mobility function~
It is known that the dependence of the mobility and the phase-field parameter $c$
produces important changes during the coarsening process \cite{PhysRevE.60.3564}. 
In the absence of studies on the appropriate mobility function for lateral phase separation in liposomes,
we consider degenerate mobility of the form $M = D c\,(1-c)$, which is a popular choice for analytical and numerical studies.
Here, $D > 0$ is a diffusivity constant. 

Parameter  $\epsilon$ in the definition of the mixture free energy defines the width of transition layer between ordered and disordered phases.  The transition layer width found in \cite{Risselada2008} is about 5 nm. Since a typical diameter of a spherical GUV in our experiments is \SI{10}{\micro\metre}, the transition layer width can be estimated as 0.1\% of its radius. This is  beyond the  capabilities of  discrete continuum model which requires the resolution of layer by the mesh. In experiments we let $\epsilon=$\SI{0.1}{\micro\metre}. This can be interpreted as letting tension forces act in a wider strip between phases, while preserving the produced momentum.   Finally, the mobility coefficient $M$ is a \emph{modeling} parameter that determines the rate of change of the order indicator  $c$ depending on the specific free energy fluctuations. In \cite{zhiliakov2021experimental}, we estimated $D$ to be in the range of $10^{-5}(\mbox{cm})^2$s$^{-1}-2.5\,10^{-5}(\mbox{cm})^2$s$^{-1}$  depending on membrane  composition. For numerical simulations presented in Sec.~\ref{sec:res}, we set $D=10^{-5}(\mbox{cm})^2$s$^{-1}$.

In order to model an initially homogenous liposome, the surface fraction $c_0$
is defined  as a realization of Bernoulli random variable~$c_\text{rand} \sim \text{Bernoulli}(a_{\text{ld}})$
with mean value $a_{\text{ld}}$, where $a_{\text{ld}}$ denotes lipid domain area fraction,
i.e.~we set:
\begin{equation}\label{raftIC}
	c_0 \coloneqq c_\text{rand}(\vect x)\quad\text{for active mesh nodes $\vect x$}.
\end{equation}
Following the thermodynamic principles described in our previous work \cite{zhiliakov2021experimental},
we set $a_{\text{ld}}=0.71$ for the 1:2:25\% (DOPC\,:\,DPPC\,:\,Chol) composition and $a_{\text{ld}}=0.29$ for the 1:1:15\%
composition. %\anna{Are those values indeed the ones used for the simulations?} \MO{0.7:0.3 was used.}

\subsubsection{Numerical method and input data}\label{sec:num_meth}

In order to find solutions to the NSCH problem in a generic setting, one has to resort to numerical methods. 
We discretize problem~\eqref{grache-1m}-\eqref{gracke-4} with
the trace finite element method (Trace FEM), a state-of-the-art computational technique for
systems of partial differential equations (PDEs) posed on surfaces~\cite{olshanskii2017trace2}.
The first step in the application of Trace FEM is common to other finite element methods. One writes 
an equivalent integral form of the PDE system, also known in PDE theory (see, e.g.~\cite{evans2010partial})
as weak formulation. The weak formulation is obtained by multiplying the equations by smooth test functions,
integrating the equations over $\Gamma$, and applying the surface Stokes formula. 
See \cite{Palzhanov2021} for the weak form of a reformulation of problem~\eqref{grache-1m}-\eqref{gracke-4}, 
that uses auxiliary pressure $\tilde p = p+ \sigma_\gamma (f(c)-c\mu)$, with $f(c)=f_0(c)+\epsilon^2|\nabla_\Gamma c|^2$, and modifies the surface tension force accordingly for numerical purposes. 
%\anna{Do we} \anna{want to show the weak formulation or should we just refer to \cite{Palzhanov2021}?} {\color{blue} It is probably sufficient to refer to\cite{Palzhanov2021}
%and mention that numerical implementation uses  auxiliary pressure $\tilde p = p+ \sigma_\gamma (f(c)-c\mu)$, with $f(c)=f_0(c)+\epsilon^2|\nabla_\Gamma c|^2$, and so the modified surface tension force for numerical purposes.  }
%For problem~\eqref{grache-1m}-\eqref{gracke-4}, 
%the weak formulations reads: Find $\bu$, $p$, $c$, and $\mu$ such that
%\begin{align}
%&\int_\Gamma \left( \rho \partial_t \bu\cdot\bv + \rho(\nabla_\Gamma\bu)\bu \cdot \bv + 2\eta  E_s(\bu):E_s(\bv)\right) \, ds - \int_\Gamma p\,\divG \bv \, ds =  -\int_\Gamma \sigma_\gamma  c  \gradG \mu \cdot\bv \, ds  \cl
%& \quad \quad + \int_\Gamma M (\nabla_\Gamma(\theta\bu)) (\gradG \mu) \cdot(\theta\bv)  \, ds, \label{gracke-w1} \\
%& \int_\Gamma q\,\divG \bu \, ds = 0, \label{gracke-w2} \\
%&\int_\Gamma \partial_t c \,v \, ds - \int_\Gamma c \bu \cdot  \gradG v \, ds + \int_\Gamma M \gradG \mu \cdot \gradG v \, ds = 0, \label{gracke-w3} \\
%&\int_\Gamma  \mu \,g \, ds =  \int_\Gamma \frac{1}{\epsilon} f_0'(c) \,g \, ds + \int_\Gamma \epsilon \gradG c \cdot \gradG g \, ds, \label{gracke-w4}
%\end{align}
%for any sufficiently regular test functions $\bv$, $q$,  $v$ and $g$ on $\Gamma$.

\begin{wrapfigure}{l}{0.24\textwidth}
\centering
\begin{overpic}[width=.24\textwidth, grid=false]{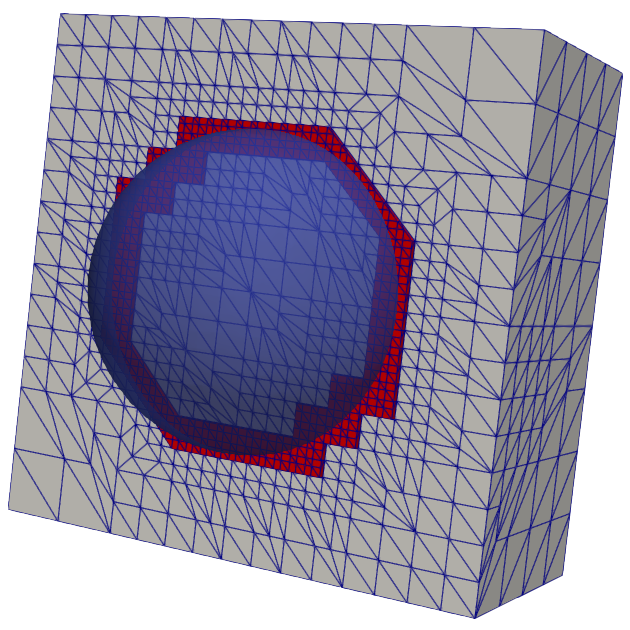}
\end{overpic}
\caption{\small A model liposome (blue) immersed in a bulk tetrahedral mesh (gray) coarser
than the actual computational mesh for visualization purposes.}
\label{fig:grid}
\end{wrapfigure}

The remaining steps are specific to Trace FEM.
The sphere $\Gamma$ is immersed in a cube, which is meshed into tetrahedra \cite{ciarlet2002finite}.
See Fig.~\ref{fig:grid}. % \anna{(Yerbol: figure to be updated not to incur in copyright issues)}.
The zero level set of the $P_1$ (i.e., first order polynomial) Lagrangian interpolant (to the vertices of the tetrahedra) of the signed 
distance function of $\Gamma$ provides a polyhedral approximation $\Gamma_h$ of the sphere, which will further be used for numerical integration in the weak formulation instead of $\Gamma$. Tetrahedra intersected
by $\Gamma_h$ form an active mesh $\mathcal{T}^{\rm bulk}$ that supports the degrees of freedom
(red layer in Fig.~\ref{fig:grid}). On  $\mathcal{T}^{\rm bulk}$ we further define finite 
dimensional spaces of continuous functions, which are polynomials of degree~either 1 or 2 on each tetrahedra 
of $\mathcal{T}^{\rm bulk}$. 
Let the restrictions of $\bu_h$, $p_h$, $c_h$, and $\mu_h$ to  $\Gamma_h$ be the approximations of $\bu$, $p$, $c$, and $\mu$. Replacing $\bu$, $p$, $c$, $\mu$ in the integral formulation 
with $\bu_h$, $p_h$, $c_h$, $\mu_h$ and choosing the test functions in the corresponding spaces 
leads to a large (but finite) system of ordinary differential equations (ODEs). Details of the  Trace FEM for the NSCH equations on surfaces are given in \cite{Yushutin_IJNMBE2019,Palzhanov2021}.
The fidelity of the numerical solution is ensured by a sequence of mesh
refinements until the solutions on two subsequent  meshes demonstrate the same
qualitative and quantitative behavior~\cite{ORG09,elliott1992error}. 
For the results in this paper we adopted mesh with 225822 active degrees of freedom (193086 for $\bu_h$
and 10912 for $p_h$, $c_h$, and $\mu_h$).
In order to contain the computational time required by the solution of the system of ODEs, we use time-stepping scheme that
(i) decouples the fluid and phase-field equation solvers at each time step \cite{Palzhanov2021} and (ii) makes use of an adaptive time stepping
technique~\cite{gomez2008isogeometric}. The time step $\Delta t$ adaptively varies
from  $\Delta t=$4$\times 10^{-6}$s during the fast initial phase of spinodal decomposition to about $\Delta t=$8$\times 10^{-4}$s
during the later slow phase of lipid domain coarsening and growth, and up to $\Delta t= 4$s when the process is close to equilibrium.
%\MO{Yerbol: what was the restriction on the max. time step?}
The piecewise polynomial approximations $\bu_h$, $p_h$, $c_h$, and $\mu_h$ are computed at every
time $t^{n+1} = t^{n} + \Delta t$ till $t^{\rm final}=4000$ s. 
%\anna{Do we want to give details about the linear algebra solver?} \MO{No!}

Just like the method proposed in \cite{zhiliakov2021experimental}, the finite element method described 
here produces numerical solutions that satisfy the mass conservation principle behind \eqref{grache-1m}-\eqref{gracke-4}:
\begin{equation}\label{raftFracDiscrete}
\int_{\Gamma_h} c_h(\vect x, t_n) \diff{s}=\int_{\Gamma_h} c_h(\vect x, t_{n-1}) \diff{s}\quad\text{implying}\quad	\frac{\int_{\Gamma_h} c_h(\vect x, t_n) \diff{s}}{\int_{\Gamma_h} 1 \diff{s}} \simeq a_{\text{ld}}
\end{equation}
for all $n=1,\dots,N$. Following \cite{zhiliakov2021experimental}, we also consider the total perimeter of the lipid domains $p_{\text{ld}}$
as a quantity of interest for the quantitative comparison with the experiments. We remark that numerically $p_{\text{ld}}$ is computed as
\begin{equation}\label{perimeter}
p_{\text{ld}}(t_n):=2\pi\int_{\Gamma_h}\epsilon|\nabla_\Gamma c_h(\vect x, t_n)|^2 \diff{s}.
\end{equation}
See \cite{zhiliakov2021experimental} for more details on this.

\section{Results and Discussion}\label{sec:res}

We focused on a ternary membrane composition DOPC:DPPC:Chol, which is known to separate
into co-existing liquid ordered ($l_o$) and liquid disordered ($l_d$) phases near room temperature
when mixed in proper ratios \cite{PMID:20642452,veatch2003separation}. Upon phase separation, 
the $l_d$ phase is composed primarily of DOPC and the $l_o$ phase is primarily composed of Chol and DPPC.
The relative size of these two phases can be tuned by adjusting the molar ratio of the lipid components. 
To assess our model of phase separation, we decided to focus our experiments on two membrane compositions 
that provide distinct and nearly opposite phase-behavior: one composition with majority $l_o$ phase and the other one with a minority 
$l_o$ phase.
We chose membranes composed of DOPC:DPPC:Chol at molar ratio of 1:1:15\%, in which the $l_o$ phase is predicted 
to occupy about 29\% of the membrane surface at 25\textdegree{}C and 1:2:25\%, in which the $l_o$ phase would occupy about 
70\% of the membrane area at 15\textdegree{}C. These area fractions were calculated using an approach 
described in our previous study \cite{zhiliakov2021experimental}  that relies on the composition of each phase 
(determined based on the phase diagram tie-lines) and the molecular area of the lipid components. 
The Cahn--Hilliard model used in \cite{zhiliakov2021experimental}, as well as continuum based models applied in other  studies~\cite{Wang2008,lowengrub2009phase,sohn2010dynamics,Li_et_al2012,Funkhouser_et_al2014}, would predict nearly the same evolution
of the domain ripening process for these two compositions since it does not account for in-membrane viscous and transport effects. 
However, the experimental data presented in this section
reveal a different domain ripening dynamics, which can be correctly captured by the more complex 
NSCH model described in Sec.~\ref{sec:math_model}.
Table \ref{tab:thermo} summarizes the lipid composition of each phase in the examined membranes along with 
the fraction of lipids in each phase and the membrane area fraction of the ordered phase for the corresponding membranes.

\begin{table}[ht]
\centering
\begin{tabular}{ |c|c|c|c|c|c|c|c|c|c|}
 \hline
 \footnotesize{Membrane composition} & \multicolumn{3}{| c |}{\footnotesize{Liquid ordered ($l_o$)}}  & \multicolumn{3}{| c |}{
 \footnotesize{Liquid disordered ($l_d$)}} &  \multicolumn{2}{| c |}{\footnotesize{Lipid fraction}} & \footnotesize{Area fraction} \\
 \hline
 \footnotesize{DOPC:DPPC:Chol (Temp)} & \footnotesize{DOPC} & \footnotesize{DPPC} & \footnotesize{Chol} & \footnotesize{DOPC} & \footnotesize{DPPC} & \footnotesize{Chol} & \footnotesize{$\alpha_{l_o}$} & \footnotesize{$\alpha_{l_d}$} & \footnotesize{$a_{\text{ld}}$}\\
  \hline
 \footnotesize{1:1:15\% (25\textdegree{}C)} & \footnotesize{16\%} & \footnotesize{58\%} & \footnotesize{26\%} & \footnotesize{57\%} & \footnotesize{34\%} & \footnotesize{9\%} & \footnotesize{0.35} & \footnotesize{0.65} & \footnotesize{29.4}\% \\
 \hline
 \footnotesize{1:2:25\% (15\textdegree{}C)} & \footnotesize{12\%} & \footnotesize{58\%} & \footnotesize{30\%} & \footnotesize{70\%} & \footnotesize{22\%} & \footnotesize{8\%} & \footnotesize{0.78} & \footnotesize{0.22} & \footnotesize{70.4\%} \\
  \hline
\end{tabular}
\caption{The composition of lipids in the liquid ordered and disordered phases (based on phase diagram), 
fraction of lipids in each phase, and the estimated domain area fraction for the two examined membrane compositions.}\label{tab:thermo}
\end{table}%

\begin{figure}[htb]
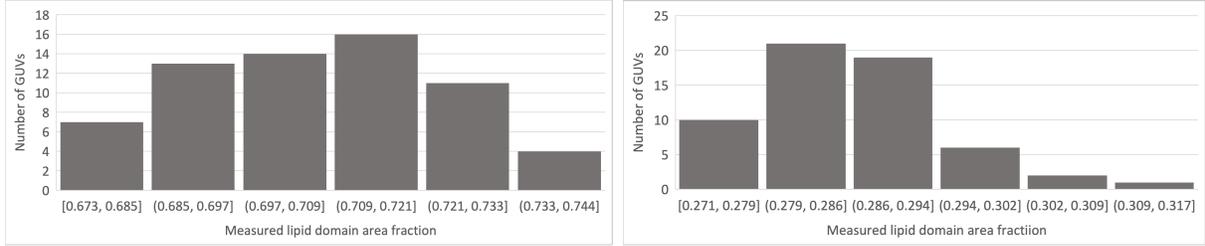

\centering
	\begin{subfigure}{.49\textwidth}
		\centering
		\includegraphicsw{{raft_area_fraction_121}.png}
	\end{subfigure}
		\begin{subfigure}{.47\textwidth}
		\centering
		\includegraphicsw{{raft_area_fraction_1115}.png}
	\end{subfigure}%
	\caption{Left: Distribution of experimental measurements of the lipid domain area fraction, with average 0.708 and standard deviation 0.017,
	 for composition 1:2:25\%. The total number of measurements is 65 and they are related to 20 GUVs.
	 Right: Distribution of experimental measurements of the lipid domain area fraction with average 0.287 and standard deviation 0.008, for
	composition 1:1:15\%. The total number of measurements is 59 and they are related to 18 GUVs.}
	\label{fig:raft_area}		
\end{figure}

\begin{figure}[htb]
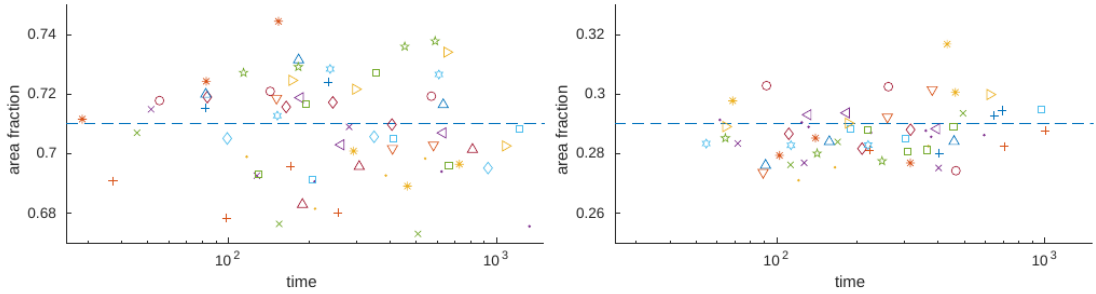

		\centering
	\begin{subfigure}{.92\textwidth}
		\includegraphicsw{{area_fraction_in_time}.png}
	\end{subfigure}%
%\begin{subfigure}{.49\textwidth}
%		\centering
%		\includegraphicsw{{area_fraction_16_rev}.png}
%	\end{subfigure}%	
	\caption{Experimentally measured lipid domain area fraction over time for composition 1:2:25\% (left) and 1:1:15\% (right). Different markers correspond to different GUVs and the dashed line represents the linear fit to the experimentally measured lipid domain area fractions.}
	\label{figS}		
\end{figure}

\begin{figure}[htb]
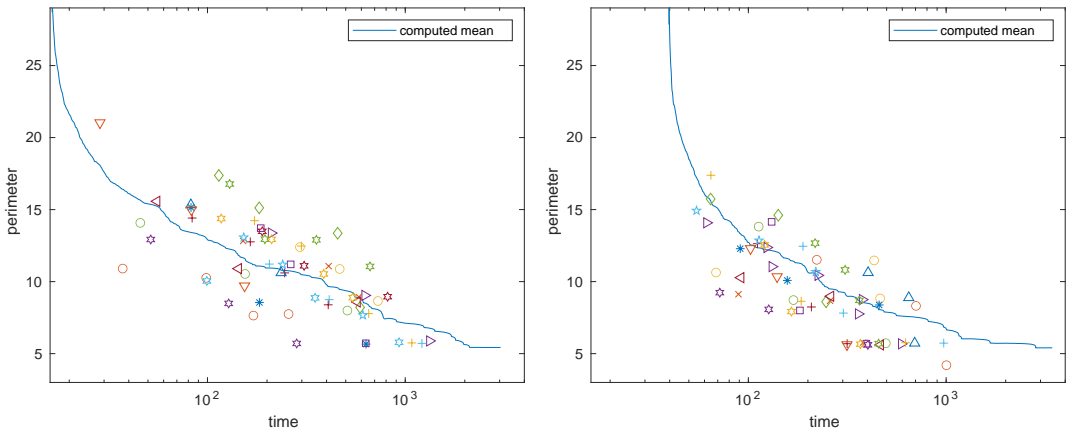

\centering
	\begin{subfigure}{.42\textwidth}
		\centering
		\includegraphicsw{{PE71_nofit}.pdf}
	\end{subfigure}~
		\begin{subfigure}{.42\textwidth}
		\centering
		\includegraphicsw{{PE29_nofit}.pdf}
	\end{subfigure}%
	\caption{Total lipid domain perimeter in \SI{}{\micro\metre} over time for composition 1:2:25\% (left)
	 and 1:1:15\% (right): numerical results average (solid line) and experimental data (markers).}
	\label{fig:perimeter}		
\end{figure}

\begin{figure}[htb]
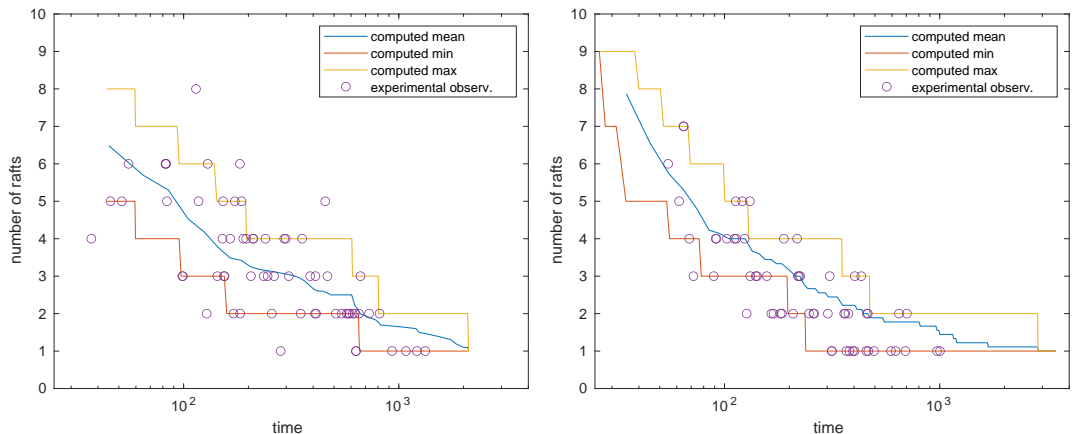

\centering
	\begin{subfigure}{.42\textwidth}
		\centering
		\includegraphicsw{{raftcount71}.pdf}
	\end{subfigure}~
\begin{subfigure}{.42\textwidth}
		\centering
		\includegraphicsw{{raftcount29}.pdf}
	\end{subfigure}%	
	\caption{Total number of lipid domains
	over time for composition 1:2:25\% (left) and 1:1:15\% (right): numerical results average (solid blue line),
	 minimum and maximum values found numerically (solid orange and yellow lines, respectively), and experimental data (circles). }
	\label{fig:Raft}		
\end{figure}

\begin{figure}[htb!]
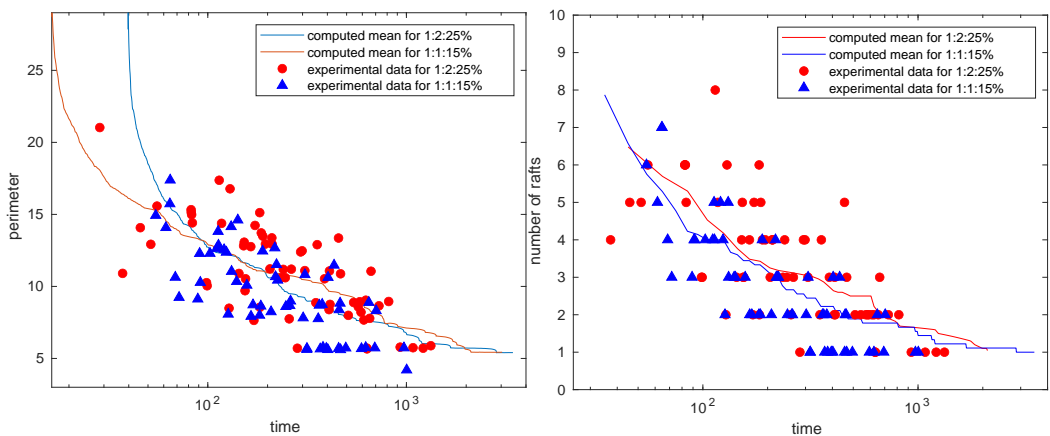

		\centering
	\begin{subfigure}{.42\textwidth}
		\includegraphicsw{{PE_overlay_new}.pdf}
	\end{subfigure}%
\begin{subfigure}{.42\textwidth}
		\centering
		\includegraphicsw{{raftcount_overlay_new}.pdf}
	\end{subfigure}%	
	\caption{Superimposition of experimental data for composition 1:2:25\% (red dots) and 1:1:15\% (blue triangles)
	with the corresponding computed means (solid line with corresponding color) for the total lipid domain perimeter (left) and
	total number of lipid domains (right).}
	\label{fig:superimposed}		
\end{figure}

\begin{figure}
\begin{center}
\href{https://youtu.be/9Zge6gqX0Ww}{
\begin{overpic}[width=.18\textwidth,grid=false]{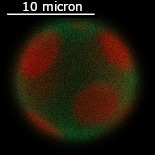}
\put(35,102){\small{$t = 102$}}
\end{overpic}~~
\begin{overpic}[width=.18\textwidth,grid=false]{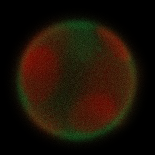}
\put(35,102){\small{$t = 145$}}
\end{overpic}~~
\begin{overpic}[width=.18\textwidth,grid=false]{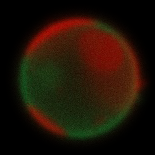}
\put(35,102){\small{$t = 408$}} 
\end{overpic}~~
\begin{overpic}[width=.18\textwidth,grid=false]{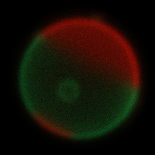}
\put(33,102){\small{$t = 1030$}}
\end{overpic}
\\
\begin{overpic}[width=.2\textwidth,grid=false]{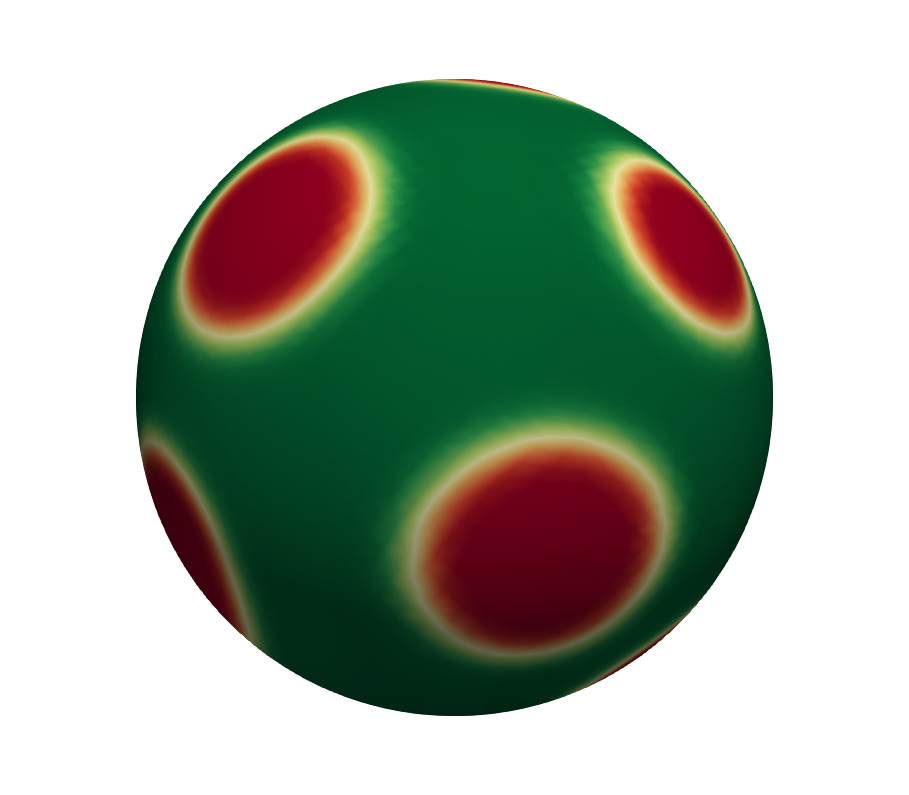}
\end{overpic}
\begin{overpic}[width=.2\textwidth,grid=false]{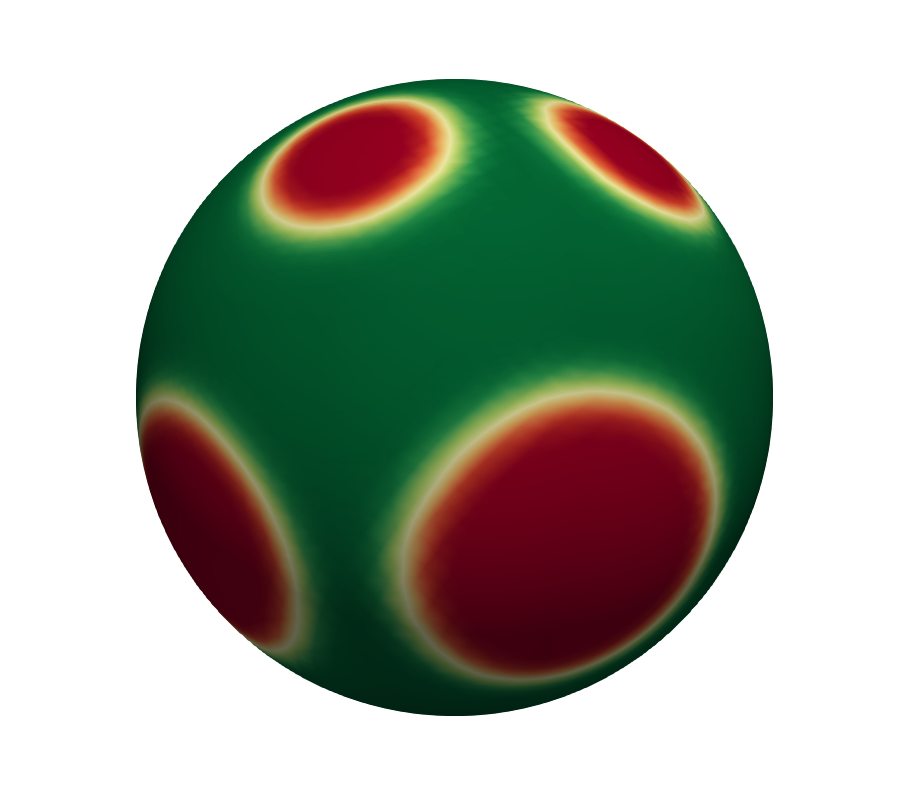}
\end{overpic}
\begin{overpic}[width=.2\textwidth,grid=false]{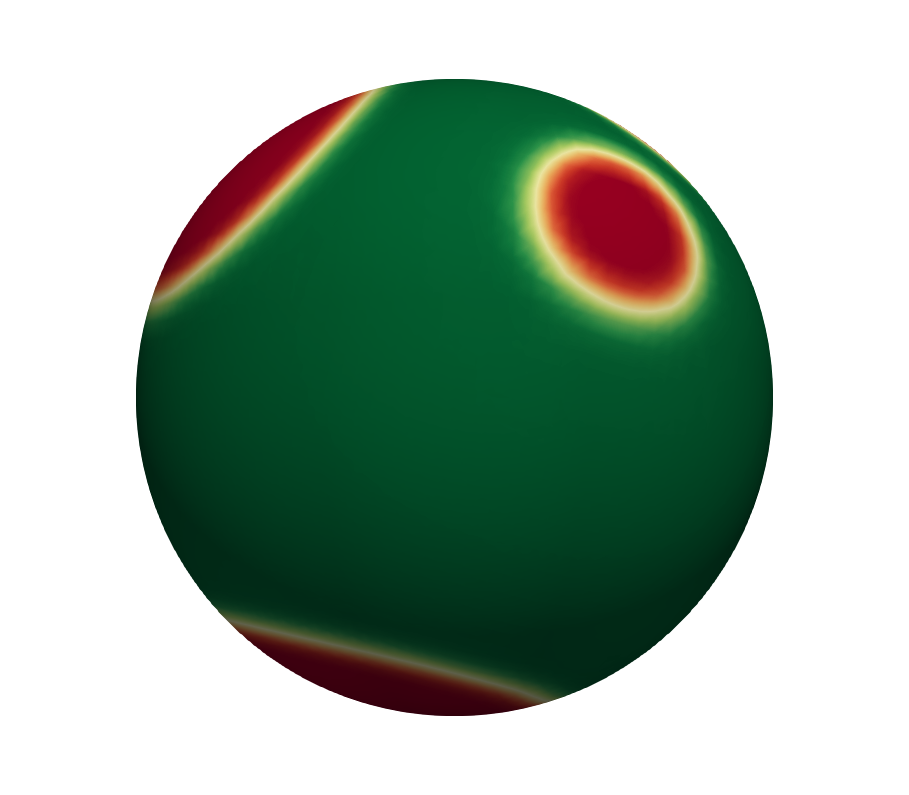}
\end{overpic}
\begin{overpic}[width=.2\textwidth,grid=false]{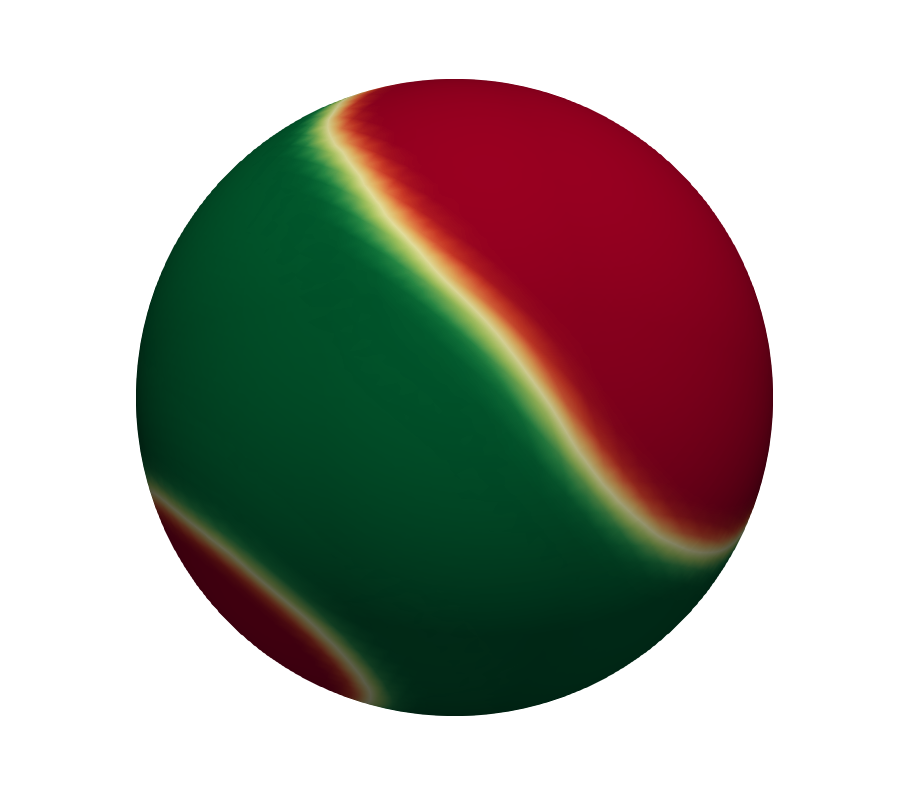}
\end{overpic}
}
\end{center}
\caption{Qualitative comparison for 1:2:25\%: epi-fluorescence microscopy images (with black background)
and numerical results (with white background) at four different times in time interval $[102, 1030]$ s. Click any picture above to run the full animation of a representative simulation.}\label{fig:qualitative_71}
\end{figure}

\begin{figure}
\begin{center}
\href{https://youtu.be/Rjyki0hZQ1U}{
\begin{overpic}[width=.16\textwidth,grid=false]{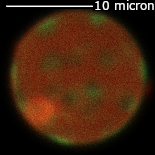}
\put(37,102){\small{$t = 73$}}
\end{overpic}~~
\begin{overpic}[width=.16\textwidth,grid=false]{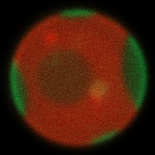}
\put(35,102){\small{$t = 166$}}
\end{overpic}~~
\begin{overpic}[width=.16\textwidth,grid=false]{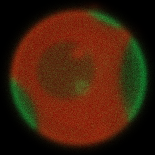}
\put(35,102){\small{$t = 225$}} 
\end{overpic}~~
\begin{overpic}[width=.16\textwidth,grid=false]{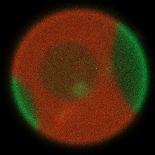}
\put(35,102){\small{$t = 244$}}
\end{overpic}~~
\begin{overpic}[width=.16\textwidth,grid=false]{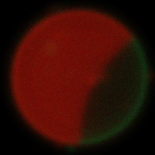}
\put(35,102){\small{$t = 322$}}
\end{overpic}
\\
\begin{overpic}[width=.15\textwidth,grid=false]{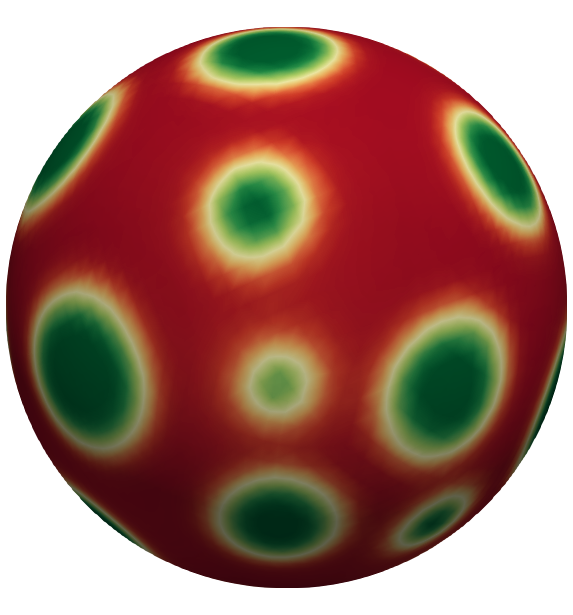}
\end{overpic}~~~
\begin{overpic}[width=.15\textwidth,grid=false]{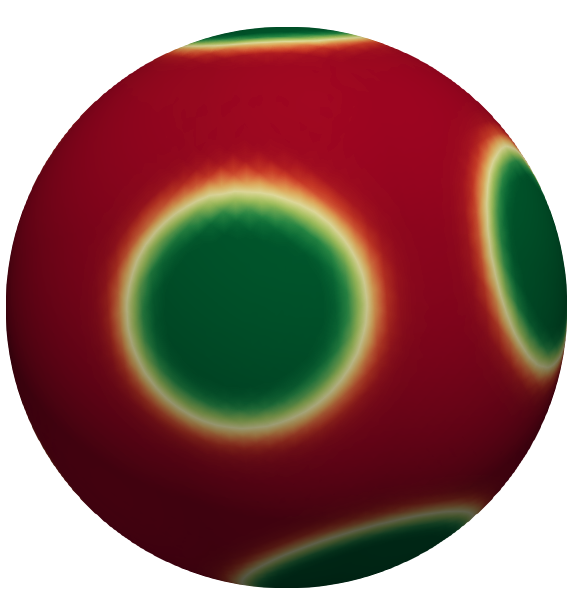}
\end{overpic}~~~~
\begin{overpic}[width=.15\textwidth,grid=false]{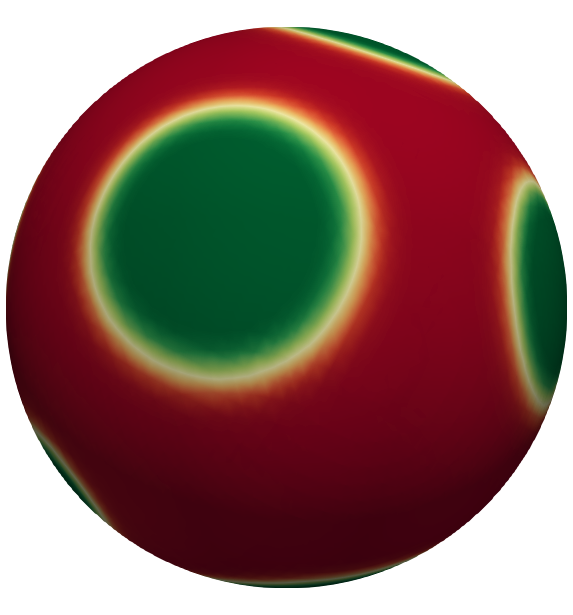}
\end{overpic}~~~~
\begin{overpic}[width=.15\textwidth,grid=false]{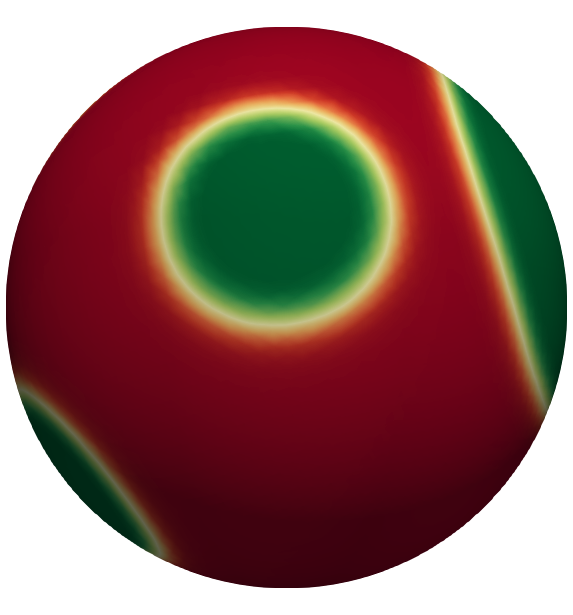}
\end{overpic}~~~
\begin{overpic}[width=.15\textwidth,grid=false]{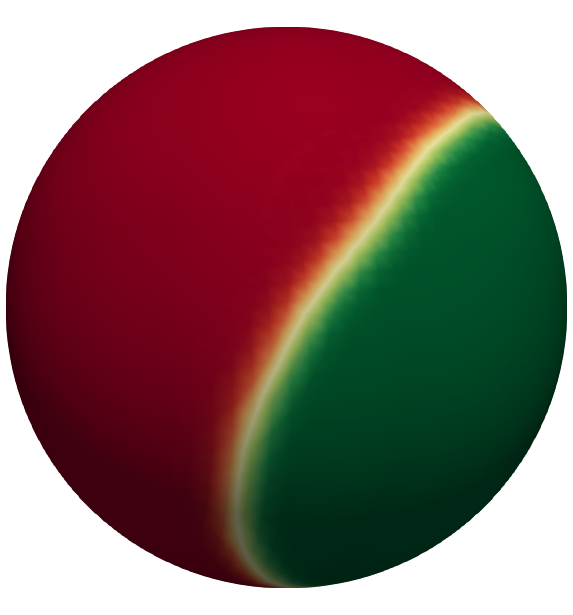}
\end{overpic}
}
\end{center}
\caption{Qualitative comparison for 1:1:15\%: epi-fluorescence microscopy images (with black background)
and numerical results (with white background) at five different times in time interval $[73, 322]$ s.
Click any picture above to run the full animation of a representative simulation.}\label{fig:qualitative_29}
\end{figure}

We applied a modified form of electroformation to produce GUVs of these two compositions and 
studied phase separation on these GUVs at temperature ranging in
[15.8, 17.5]\textdegree{}C for 1:2:25\% and [23.8,25.7]\textdegree{}C for 1:1:15\% composition.
Note that the experimental temperature ranges were selected to match the temperatures used for the above-mentioned 
theoretical calculations.  
Using confocal fluorescence microscopy, we examined a minimum of 18 GUVs 
(from 4-5 independent experiments) for the number of their lipid domains as well as area and perimeter of domains 
at different time points, for each GUV composition.
The fraction of vesicle surface area occupied by the $l_o$ phase, i.e.~lipid domain area fraction,
in GUVs was calculated from the confocal images. The results are summarized in Fig.~\ref{fig:raft_area}, which shows
the distribution of lipid domain area fractions for both GUV compositions.
%, respectively, along with their corresponding average and standard deviation.
The experimental $l_o$ domain area fractions, 0.287$\pm$0.008 and 0.708$\pm$0.017, are in great agreement with 
those predicted above (Table \ref{tab:thermo}) based on the literature-reported phase diagrams, 0.294 and 0.704, for 1:1:15\% and 
1:2:25\% compositions, respectively. This agreement validates our experimental results. 
We report in Fig.~\ref{figS} the experimentally measured lipid domain area fraction over time for the two compositions.
We observe only slight (i.e. non-significant) changes on a given GUV as time passes, i.e.~as the number
of domains reduces. This result supports one of the assumptions behind the NSCH model, that is the 
conservation of lipid domain area fraction over time. 
The time on the horizontal axis in Fig.~\ref{figS}-\ref{fig:superimposed} corresponds to the time
as measured in the experiments, as is the case for the times reported in Fig.~\ref{fig:qualitative_71}
and \ref{fig:qualitative_29}.

%To validate our experimental results, we compared these results to area fractions predicted
%by the literature-reported phase diagrams for this ternary membrane. \anna{Sheereen:  Here, we should
%summarize the results from the literature and refer to \cite{zhiliakov2021experimental} for how to calculate
%$a_{\text{ld}}$ or report the calculations here.
%}

Independent of the experimental results, 10 numerical simulations were run for each composition.
All the simulated liposomes had a \SI{10}{\micro\metre} diameter
and they differed in the realization of the random variable used to set up initial state as explained in Sec.~\ref{sec:math_model}. 
For each simulation, we tracked the total lipid domain perimeter and the total number of 
lipid domains in order to compare with the experimental data. The computed lipid domain area fraction
is not reported since, as mentioned before, the NSCH model is conservative and thus the lipid domain area fraction
stays constant over time. 

In order to compare the total lipid domain perimeter between simulations and experiments,
we first scaled all dimensional observables the depend on a length unit by the radius of the corresponding GUV, 
since the diameter of GUVs varied in the experiments
(between 9-\SI{16}{\micro\metre}) while it was constant in the simulations.
Fig.~\ref{fig:perimeter} reports all the rescaled experimental
measurements with markers (a different marker for each GUV) and the average of the computed total lipid domain perimeter
from all the simulations with a solid line for compositions 1:2:25\% and 1:1:15\%. In both cases,
the average of the computed total lipid domain perimeters falls within the cloud of experimental measurements.
We note that no experimental measurement is available before \SI{40}{\second} because no lipid domains 
were observed in this time frame,  presumably due to the small size of domains that could not be resolved under fluorescence microscopy.

Next, we performed a quantitative comparison for the total number of lipid domains on a GUV over time.
Fig.~\ref{fig:Raft} shows the experimentally measured and numerically computed data for both examined compositions.
The measurements are reported with a circle, while for the simulations we reported three solid lines corresponding
to the numerical results average, minimum, and maximum number of lipid domains found in the simulations.
We see that the vast majority of the experimental data (89\% for composition 1:2:25\% and 91\% for composition 1:1:15\%) falls within the computed extrema.

In order to facilitate the understanding of the different domain ripening dynamics
for the two membrane compositions under consideration, we superimpose the experimental data
for total lipid domain perimeter and total number of lipid domains in Fig.~\ref{fig:superimposed}.
We observe in average faster dynamics towards the equilibrium state (i.e., one domain of the minority
phase within a background of the majority phase) for composition 1:1:15\%, which has majority 
$l_d$ phase. This is correctly captured by the NSCH model described in Sec.~\ref{sec:math_model}.
Indeed, we see that the solid blue curve (corresponding to the computed mean for composition 1:1:15\%) lies below the red curve
(corresponding to the computed mean for composition 1:2:25\%) for the majority of the time interval under consideration
in both graphs in Fig.~\ref{fig:superimposed}. We recall that the simplified model in  \cite{zhiliakov2021experimental} would predict 
the same evolution for $a_{\text{ld}}=0.71$ and $a_{\text{ld}}=0.29$ and thus it would be unsuited
to reproduce the experimental data reported in this paper. 

To further compare the experimental data to the simulation results, we present a qualitative comparison between images acquired with
epi-fluorescence microscopy and images obtained from post-processing the numerical results.
Fig.~\ref{fig:qualitative_71} and Fig.~\ref{fig:qualitative_29} present such comparison for
compositions 1:2:25\% and 1:1:15\%, respectively.
Notice that the the representative microscopy images in Fig.~\ref{fig:qualitative_71} and \ref{fig:qualitative_29} refer to different
sets of GUVs than those used for the quantitative analysis in Fig.~\ref{fig:raft_area}-\ref{fig:Raft}
because confocal microscopy (needed for the measurement) and epi-fluorescence microscopy cannot be used
simultaneously. %Epi-fluorescence microscopy images could not be used for a quantitative
%comparison with the numerical simulations because they provide only a two-dimensional picture
%of the liposome. 
Overall, from Fig.~\ref{fig:qualitative_71} and \ref{fig:qualitative_29} we see an excellent
qualitative agreement between experiments and simulations.

\begin{figure}[htb!]
	\centering
	\begin{overpic}[width=.15\textwidth,grid=false]{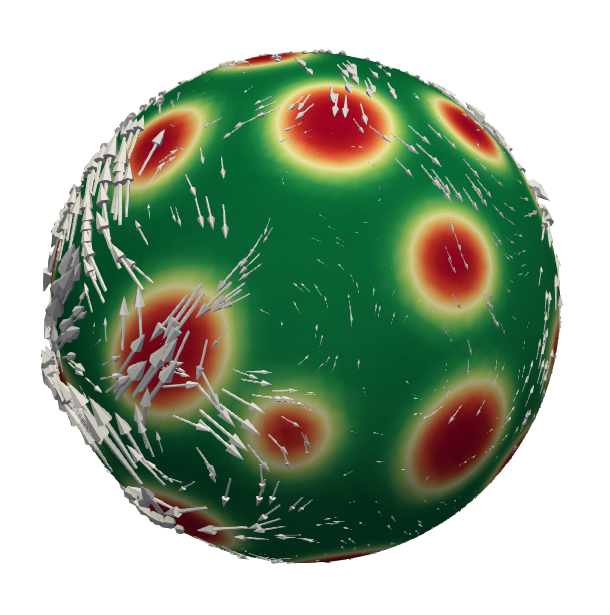}
        \put(33,98){\small{$t = 15$}}
        \put(-48,50){\small{1:2:25\%}}
        \put(-48,-5){\small{max speed:}}
        \put(30,-5){\small{0.0237}}
        \end{overpic}~~
        \begin{overpic}[width=.15\textwidth,grid=false]{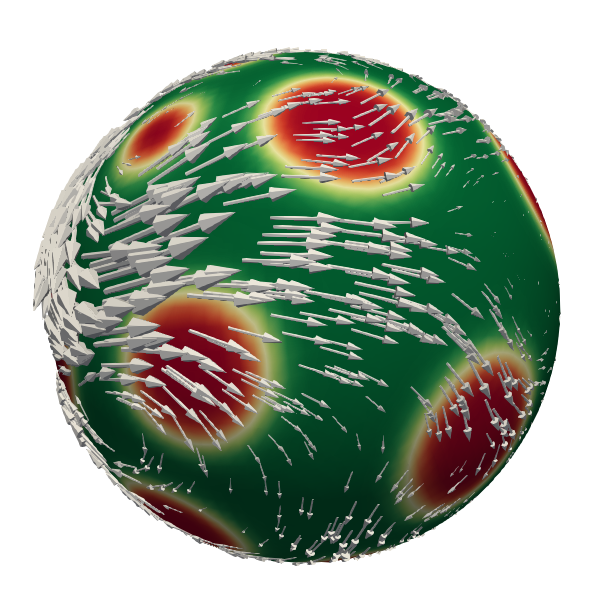}
        \put(33,98){\small{$t = 19$}}
        \put(30,-5){\small{0.0118}}
        \end{overpic}~~
        \begin{overpic}[width=.15\textwidth,grid=false]{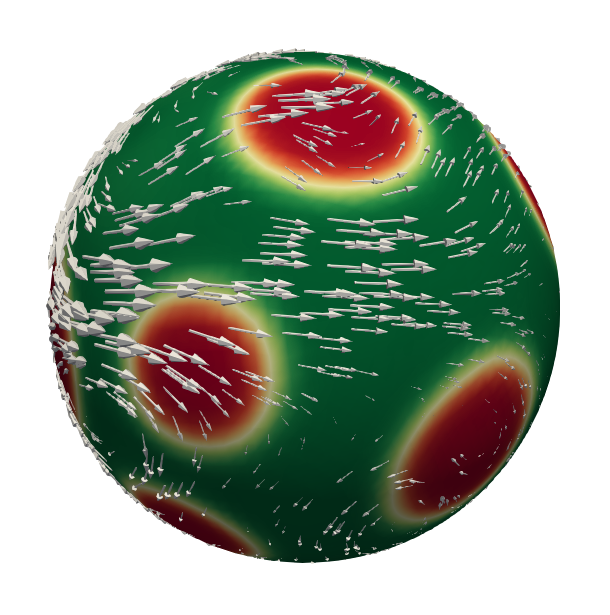}
        \put(33,98){\small{$t = 22$}}
        \put(30,-5){\small{0.0070}}
        \end{overpic}~~
        \begin{overpic}[width=.15\textwidth,grid=false]{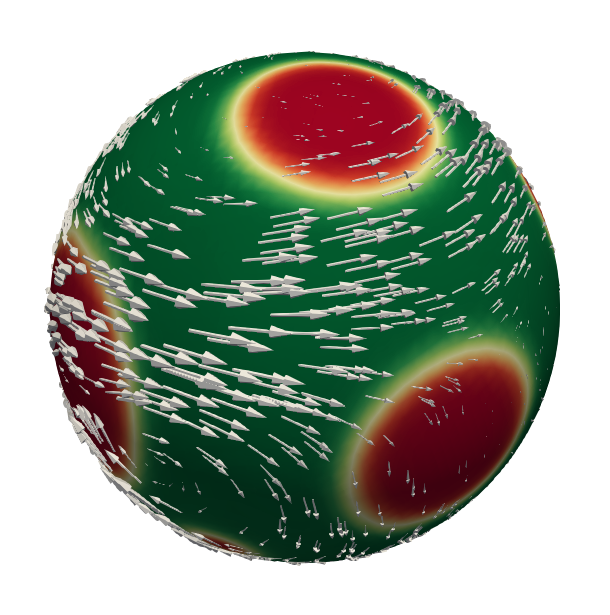}
        \put(33,98){\small{$t = 36$}}
        \put(30,-5){\small{0.0034}}
        \end{overpic}~~
        \begin{overpic}[width=.15\textwidth,grid=false]{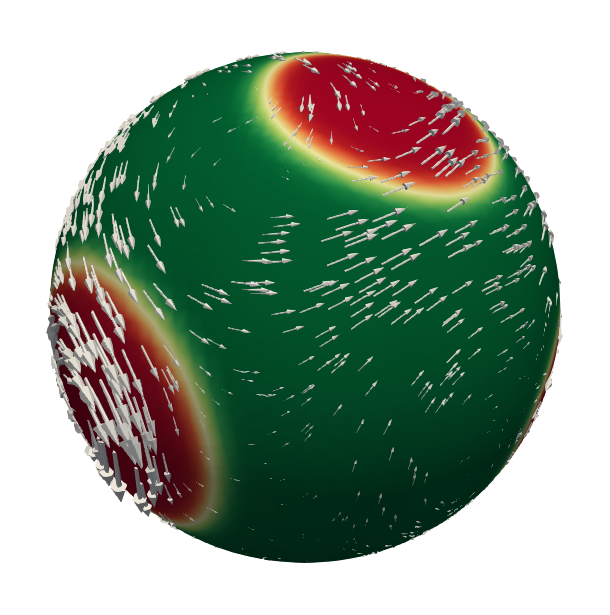}
        \put(31,98){\small{$t = 135$}}
        \put(30,-5){\small{0.0013}}
        \end{overpic}
        \\
        \vskip .5cm
	\begin{overpic}[width=.15\textwidth,grid=false]{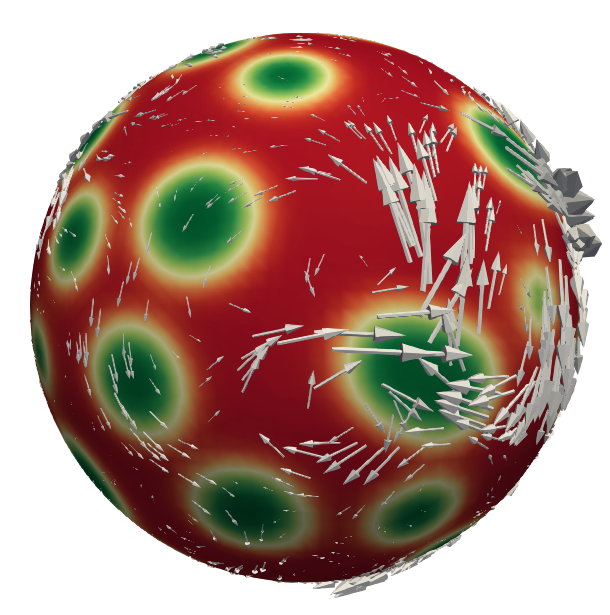}
        \put(33,98){\small{$t = 55$}}
        \put(-48,50){\small{1:1:15\%}}
        \put(-48,-7){\small{max speed:}}
        \put(30,-7){\small{0.02434}}
        \end{overpic}~~
        \begin{overpic}[width=.15\textwidth,grid=false]{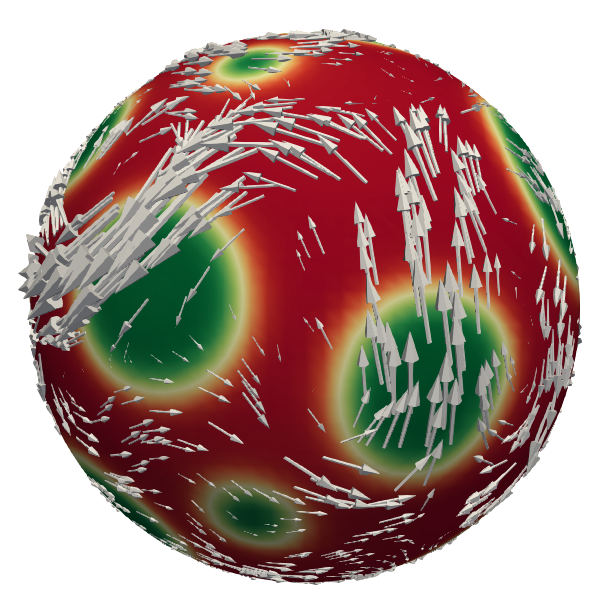}
        \put(33,98){\small{$t = 62$}}
        \put(30,-7){\small{0.0075}}
        \end{overpic}~~
        \begin{overpic}[width=.15\textwidth,grid=false]{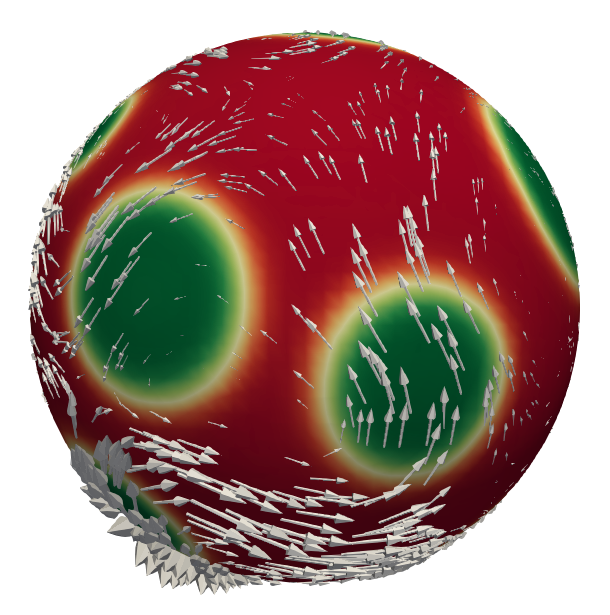}
        \put(33,98){\small{$t = 65$}}
        \put(30,-7){\small{0.0114}}
        \end{overpic}~~
        \begin{overpic}[width=.15\textwidth,grid=false]{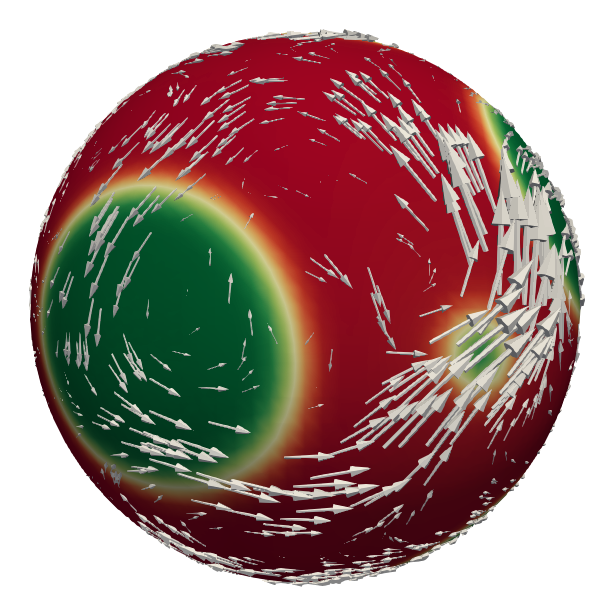}
        \put(33,98){\small{$t = 77$}}
        \put(30,-7){\small{0.0030}}
        \end{overpic}~~
        \begin{overpic}[width=.15\textwidth,grid=false]{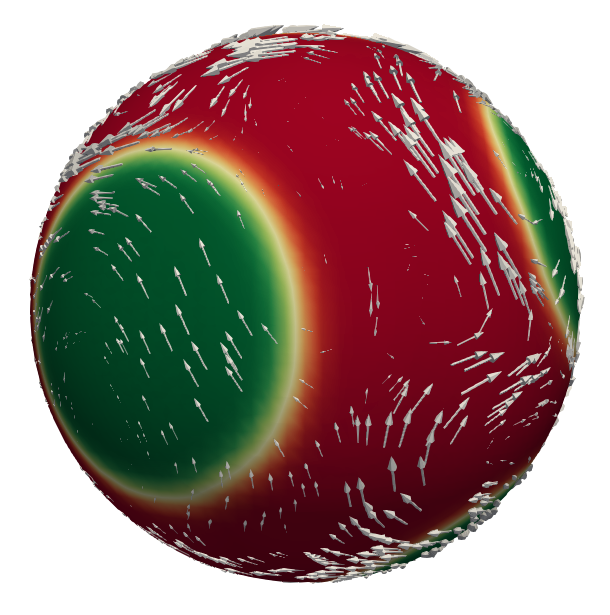}
        \put(31,98){\small{$t = 128$}}
        \put(30,-7){\small{0.0015}}
        \end{overpic}
	\caption{Composition 1:2:25\%: 
	Evolution of the velocity vectors superimposed to the lipid domains and maximum speed in \SI{}{\micro\metre}/s
	for composition 1:2:25\% (top) and 1:1:15\% (bottom). The velocity vectors are magnified by a factor
	100, 300, 300, 700, and 1500 (from left to right) for visualization purposes. 
	}\label{fig:vel_field}
\end{figure}
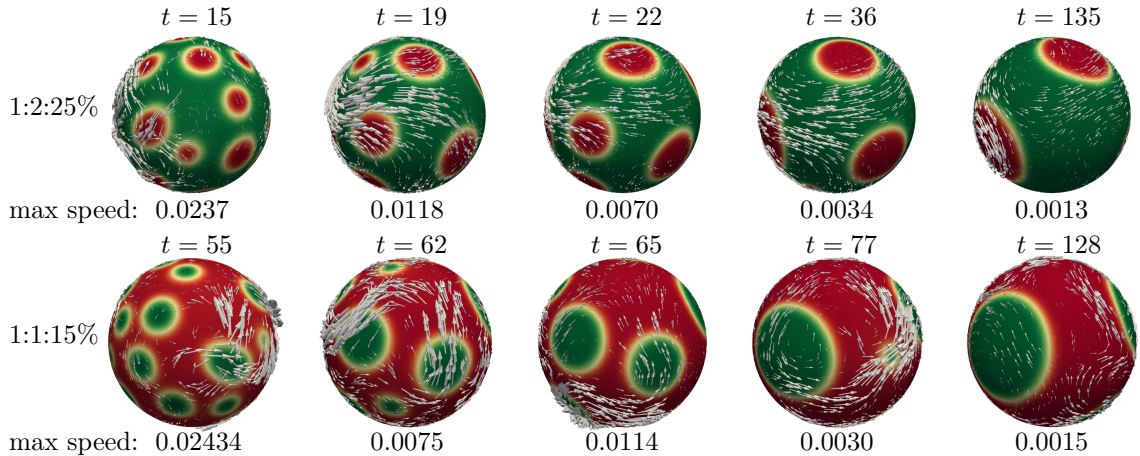

A piece of information that is provided by the numerical simulations but is impossible
to visualize or measure experimentally is the flow field. Fig.~\ref{fig:vel_field} shows the evolution of the computed velocity vectors
superimposed to the computed surface fraction for both compositions. We observe a larger 
velocity magnitude
when there are several lipid domains on the surface that are in the process of merging. The velocity magnitude
becomes smaller as the number of lipid domain decreases and the system gets closer to an equilibrium. 

\begin{wrapfigure}{r}{0.18\textwidth}
\centering
	\begin{overpic}[width=.18\textwidth,grid=false]{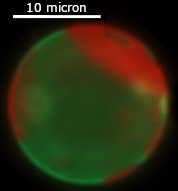}
\end{overpic}
\caption{\small Composition 1:2:25\%: combination of $l_d$ domains (red) within the $l_o$ phase (green)
and viceversa.}
\label{fig:mixed}
\end{wrapfigure}

Interestingly, in the case of composition 1:2:25\%, we observed different types of domain forma-
\begin{wrapfigure}{l}{0.18\textwidth}
\centering
	\begin{overpic}[width=.18\textwidth,grid=false]{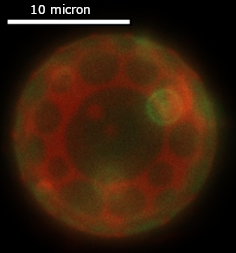}
\end{overpic}
\caption{\small Composition 1:2:25\%: formation of $l_o$ domains (green) within the $l_d$ phase (red).}
\label{fig:green_patches}
\end{wrapfigure}
tion on GUVs. 
While at equilibrium these GUVs consistently had a $l_o$ phase that occupied the majority of their surface 
(in average 70.8\%), prior to the equilibrium state, three distinct domain patterns were observed on these GUVs. 
In some GUVs, $l_d$ domains formed within the $l_o$ phase 
(i.e., red patches on green background as shown
in Fig.~\ref{fig:qualitative_71}), in others $l_o$ domains formed within the $l_d$ phase (i.e. green patches on red background, 
see Fig.~\ref{fig:green_patches}), 
and in others again a combination of these domain patterns was detected (see Fig.~\ref{fig:mixed}). 
Often, the dominating phase in the membrane forms the background 
in which domains of the minority phase 
form, merge, and grow into one larger domain at equilibrium. Domain formation in our GUVs with 1:1:15\% 
composition always followed this typical phase behavior. However, in some compositions such as 1:2:25\%, it is 
also possible that the dominating phase forms domains within the minority phase \cite{SAKUMA20201576}. 
Thus, membranes of such compositions may show either type of domains. The rather rare formation of domains 
of the dominating phase 
within the minority phase background may be due to the existence of a metastable third phase 
for such membrane compositions, which could be due to the formation of  some temporary molecular complexes.
It can be seen as an example of the Ostwald rule of stages, by which the system 
may go through a sequence of intermediate states before reaching the most energetically preferable one~\cite{van1984ostwald,anderson2002insights}. 
The metastable phase does not present as an equilibrium state (i.e., the states reported in 
phase diagrams), nonetheless it can exist long enough to influence the dynamics of the phases. 
For instance, the presence of such third phase can cause variations in domain area fraction before 
reaching a two-phase equilibrium state. In fact, we noticed that in those GUVs that formed $l_o$ domains within 
$l_d$ phase in 1:2:25\% composition, the area fraction of $l_o$ phase was not as stable and varied prior to equilibrium; 
the $l_o$ domain area fraction was on average  $\sim$6\% lower than that at equilibrium. 

\begin{figure}[htb]
\begin{center}
\href{https://youtu.be/HgHmk8vZZdE}{
\begin{overpic}[width=.15\textwidth,grid=false]{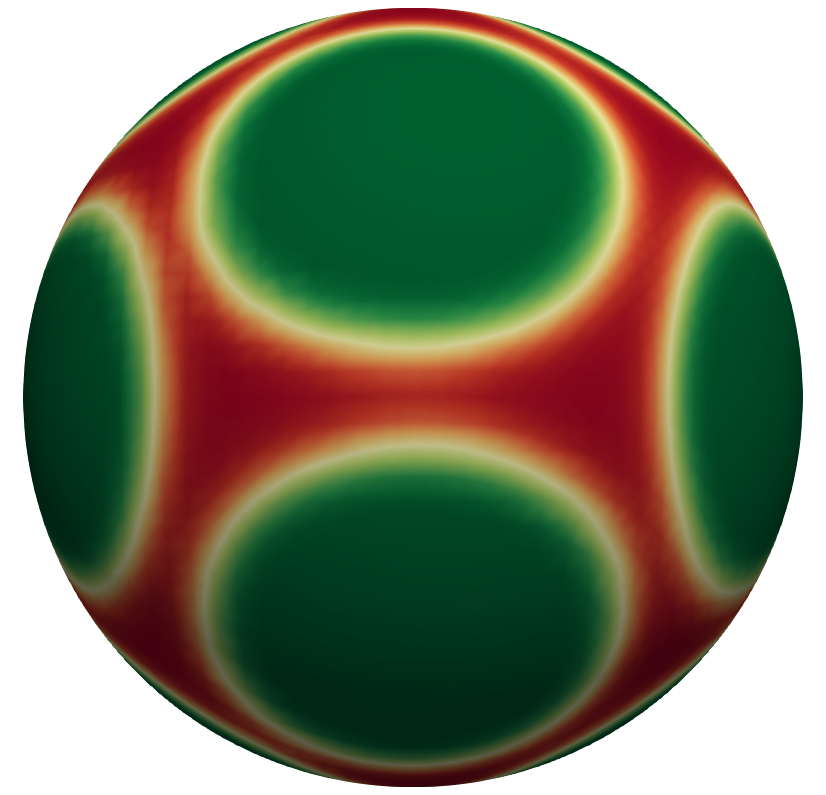}
%\put(37,102){\small{$t = 1$}}
\end{overpic}~~
\begin{overpic}[width=.15\textwidth,grid=false]{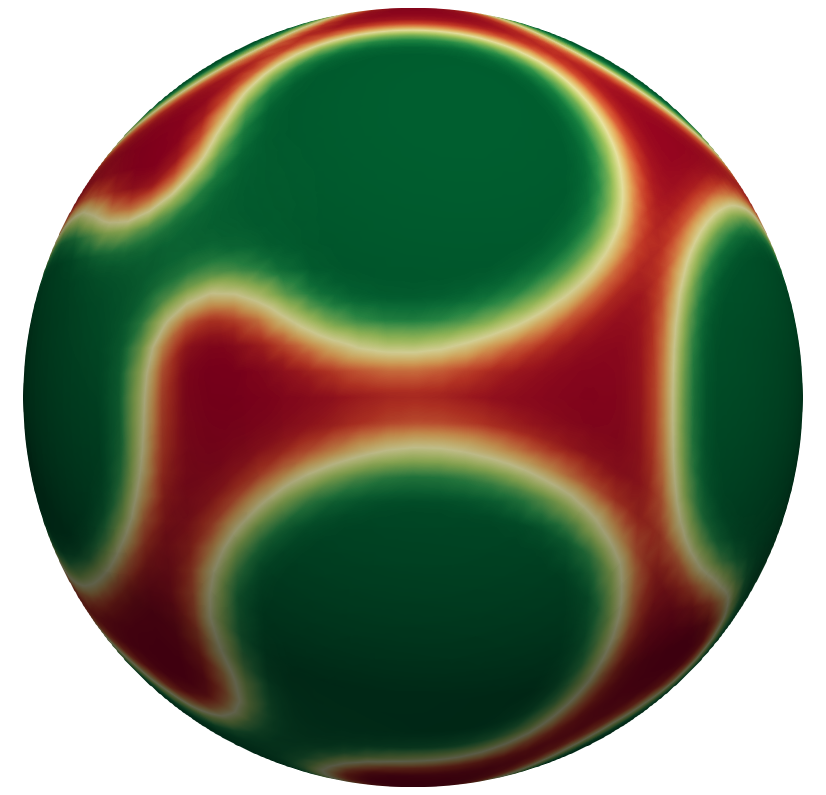}
%\put(37,102){\small{$t = 5$}}
\end{overpic}~~
\begin{overpic}[width=.15\textwidth,grid=false]{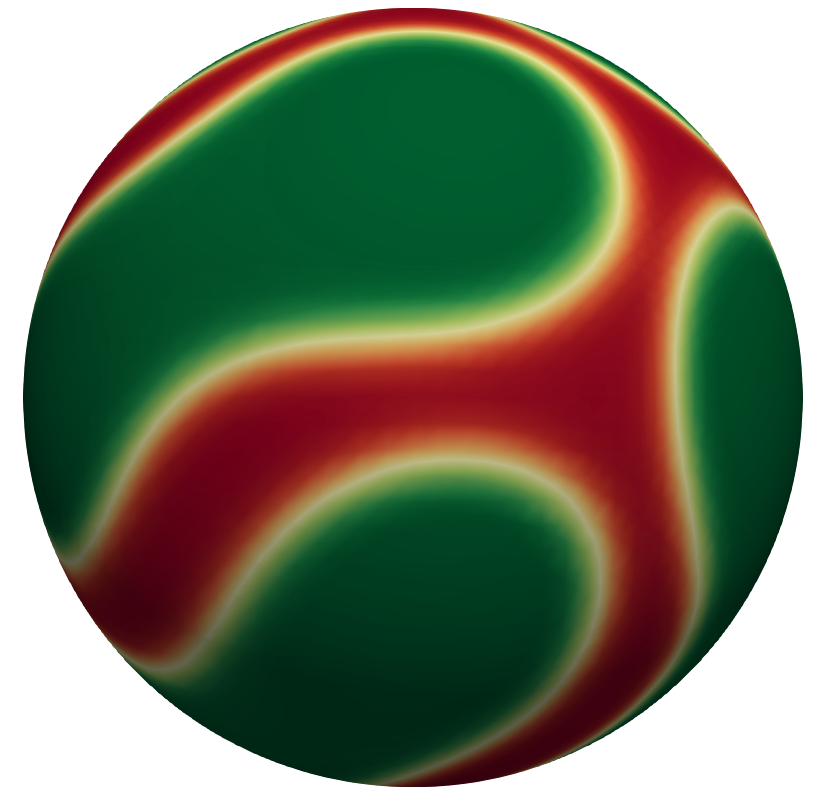}
%\put(37,102){\small{$t = 7$}}
\end{overpic}~~
\begin{overpic}[width=.15\textwidth,grid=false]{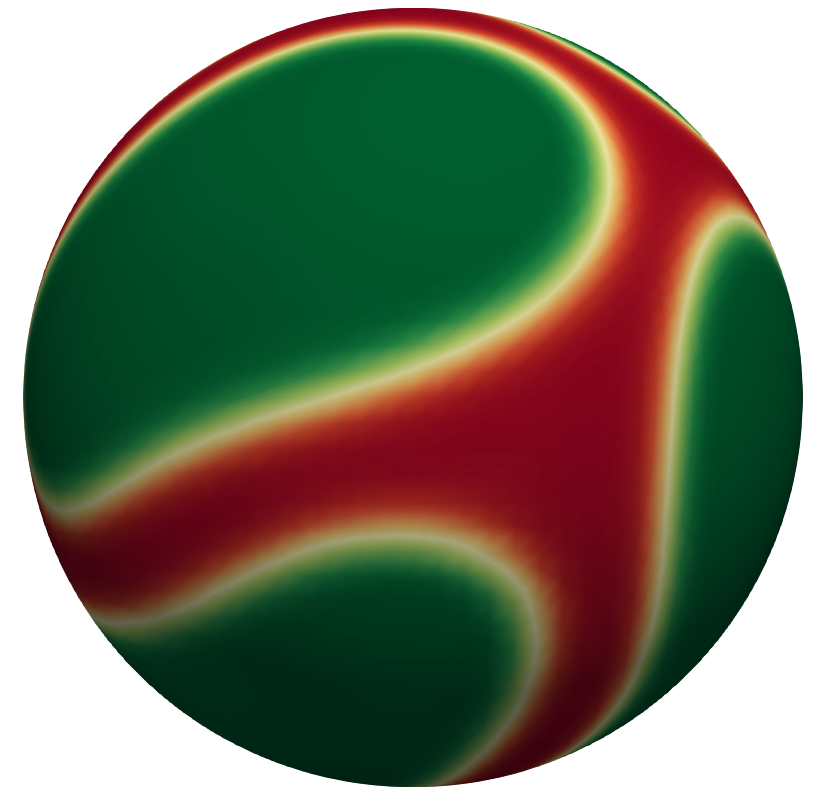}
%\put(37,102){\small{$t = 10$}}
\end{overpic}~~
\begin{overpic}[width=.15\textwidth,grid=false]{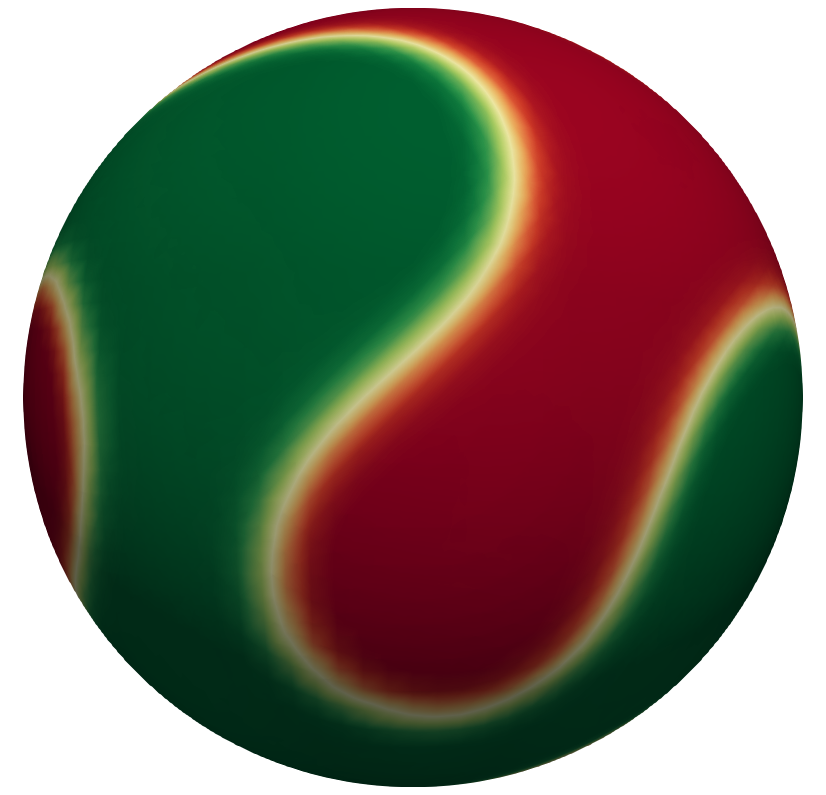}
%\put(37,102){\small{$t = ??$}}
\end{overpic}~~
\begin{overpic}[width=.15\textwidth,grid=false]{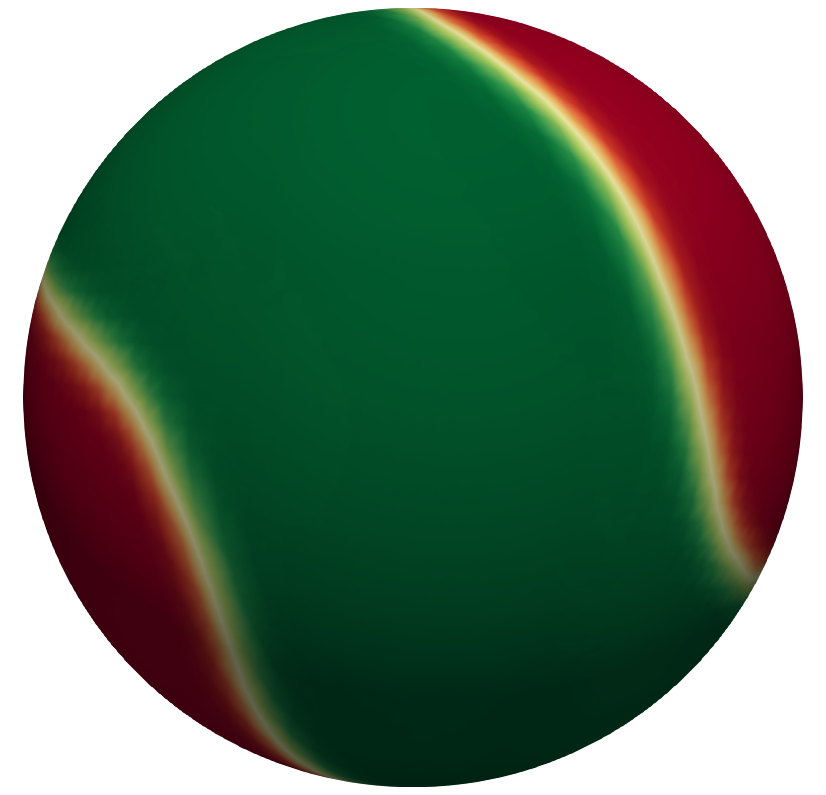}
%\put(37,102){\small{$t = ??$}}
\end{overpic}
}
\end{center}
\caption{Composition 1:2:25\%:  simulated evolution of $l_o$ domains (green) within the $l_d$ phase (red) starting 
from 12 symmetrically placed $l_o$ domains and $a_{\text{ld}}=0.64$. Time is increasing from left to right.}\label{fig:rafts_12}
\end{figure}

\begin{figure}[htb]
\begin{center}
\href{https://youtu.be/H6y32N7pb7s}{
\begin{overpic}[width=.17\textwidth,grid=false]{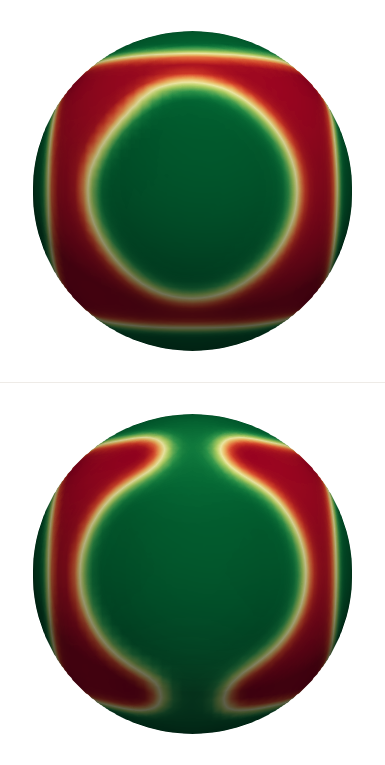}
%\put(15,102){\small{$t = 0.3$}}
\put(-30,73){\small{Front view}}
\put(-30,23){\small{Rear view}}
\end{overpic}~~
\begin{overpic}[width=.17\textwidth,grid=false]{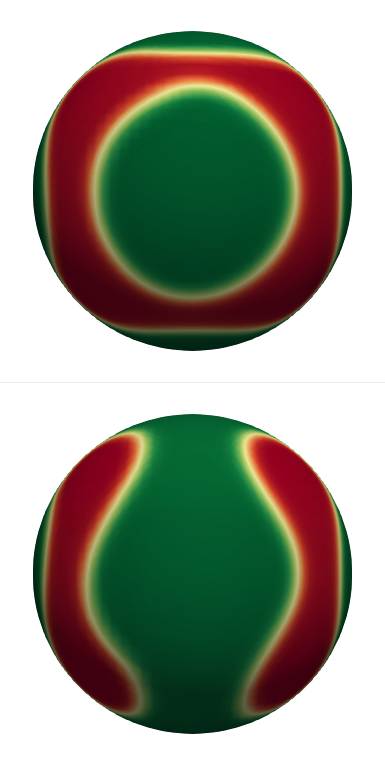}
%\put(15,102){\small{$t = 3$}}
\end{overpic}~~
\begin{overpic}[width=.17\textwidth,grid=false]{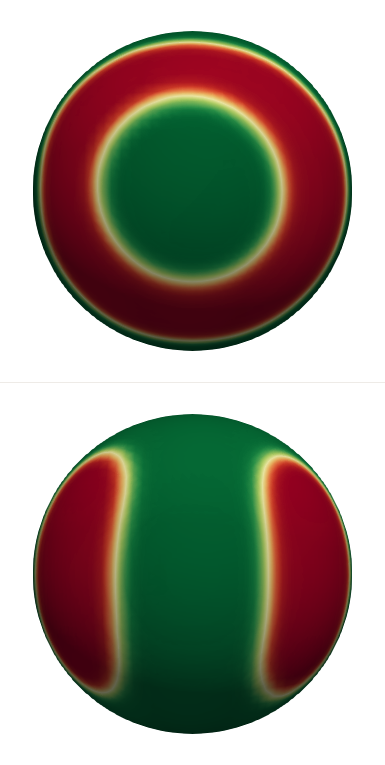}
%\put(15,102){\small{$t = 12$}}
\end{overpic}~~
\begin{overpic}[width=.17\textwidth,grid=false]{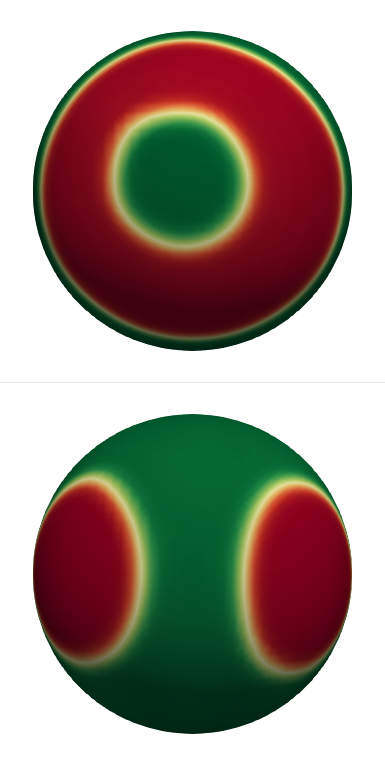}
%\put(15,102){\small{$t = 21$}}
\end{overpic}
}
\end{center}
\caption{Composition 1:2:25\%: front (top row) and rear (bottom row) view of a simulated GUV showing
mixed $l_o$ (green) and $l_d$ (red) domains starting from 6 symmetrically placed $l_o$ domains 
and $a_{\text{ld}}=0.65$. Time is increasing from left to right.
Click any picture above to run the full animation.}\label{fig:lakes_rats_6}
\end{figure}

The CHNS model \eqref{grache-1m}-\eqref{gracke-4} is not suitable to model the dynamic of a third temporary phase. 
Even reproducing the reverse phase pattern in the numerical simulations proved to be challenging as the initial conditions formulated 
in Sec.~\ref{sec:math_model} would inevitably lead to $l_d$ domains 
formed within the $l_o$ phase for composition 1:2:25\%. In an effort to emulate the effect of the metastable phase on the patter evolution, 
we introduced the following 2 changes with respect to the set up described in 
Sec.~\ref{sec:math_model}: i) the simulations are started with a number of preformed 
$l_o$ domains placed symmetrically in $l_d$ background and ii) the area fraction of $l_o$ phase is reduced by about 
6\% to match experimental evidence. Fig.~\ref{fig:rafts_12} shows the formation of 
$l_o$ domains within the $l_d$ phase starting from 12 symmetrically placed $l_o$ domains 
and $a_{\text{ld}}=0.64$. The third scenario (co-existence of $l_o$ and $l_d$ domains) is shown in
Fig.~\ref{fig:lakes_rats_6} and was obtained starting from 6 symmetrically placed $l_o$ domains 
and $a_{\text{ld}}=0.65$. 
Because of the forced initial condition described at the above point i), which differ from the 
initial composition in the experiments (i.e., a initially homogenous GUV), we preferred not to report time
in Fig.~\ref{fig:rafts_12} and \ref{fig:lakes_rats_6}.

\section{Conclusion}
This paper presents an experimental and computational study on the evolution of lipid rafts in
membranes with ternary membrane composition DOPC:DPPC: Chol. We focused on 
two specific compositions that yield opposite and nearly inverse phase behavior.
A non-equilibrium thermodynamics model describing phase separation alone, like the Cahn--Hilliard model, 
would predict nearly the same evolution of the domain ripening process for these two compositions. 
However, our experimental data collected using fluorescence microscopy reveal different domain ripening dynamics.
In order to capture this phenomenon, we considered the more complex Navier--Stokes--Cahn--Hilliard model, 
which accounts for in-membrane viscous and transport effects. The computational results obtained with this 
improved model are in excellent agreement with the experimental data in terms of domain area fraction,
total domain perimeter over time and total number of domains over time for both compositions
under consideration.

\section*{Acknowledgments}
This work was partially supported by US National Science Foundation (NSF) through grant DMS-1953535.
M.O.~acknowledges the support from NSF through DMS-2011444.
S.M.~acknowledges the support from NSF through DMR-1753328.
A.Q.~acknowledges support from the Radcliffe Institute for Advanced Study at Harvard University where
she has been a 2021-2022 William and Flora Hewlett Foundation Fellow.
The authors are grateful to Vassiliy Lubchenko for insightful discussions on thermodynamics of polymorphs
and to Alexander Zhiliakov for helping with the set up of the numerical experiments.

% Uncomment if using bibtex (default)
\bibliographystyle{plain}
\bibliography{literatur}

\begin{thebibliography}{10}

\bibitem{Abels2012}
H.~Abels, H.~Garcke, and G.~Gr\"{u}n.
\newblock Thermodynamically consistent, frame indifferent diffuse interface
  models for incompressible two-phase flows with different densities.
\newblock {\em Mathematical Models and Methods in Applied Sciences},
  22(03):1150013, 2012.

\bibitem{anderson2002insights}
Valerie~J Anderson and Henk~NW Lekkerkerker.
\newblock Insights into phase transition kinetics from colloid science.
\newblock {\em Nature}, 416(6883):811--815, 2002.

\bibitem{Bandekar2012}
A.~Bandekar and S.~Sofou.
\newblock Floret-shaped solid domains on giant fluid lipid vesicles induced by
  {pH}.
\newblock {\em Langmuir}, 28:4113--4122, 2012.

\bibitem{Bandekar2013}
Amey Bandekar, Charles Zhu, Ana Gomez, Monica~Zofia Menzenski, Michelle
  Sempkowski, and Stavroula Sofou.
\newblock Masking and triggered unmasking of targeting ligands on liposomal
  chemotherapy selectively suppress tumor growth in vivo.
\newblock {\em Molecular Pharmaceutics}, 10(1):152--160, 2013.

\bibitem{BENNETT20131765}
W.F.~Drew Bennett and D.~Peter Tieleman.
\newblock Computer simulations of lipid membrane domains.
\newblock {\em Biochimica et Biophysica Acta (BBA) - Biomembranes},
  1828(8):1765--1776, 2013.

\bibitem{brown2000}
D.A. Brown and E.~London.
\newblock Structure and function of sphingolipid- and cholesterol-rich membrane
  rafts.
\newblock {\em J Biol Chem.}, 275(23):17221--17224, 2000.

\bibitem{Brown1}
R.E. Brown.
\newblock Sphingolipid organization in biomembranes: what physical studies of
  model membranes reveal.
\newblock {\em Journal of Cell Science}, 111(1):1--9, 1998.

\bibitem{ciarlet2002finite}
Philippe~G Ciarlet.
\newblock {\em The finite element method for elliptic problems}, volume~40.
\newblock {SIAM}, 2002.

\bibitem{elliott1992error}
Charles~M Elliott and Stig Larsson.
\newblock Error estimates with smooth and nonsmooth data for a finite element
  method for the {Cahn--Hilliard} equation.
\newblock {\em Mathematics of Computation}, 58(198):603--630, 1992.

\bibitem{evans2010partial}
Lawrence~C Evans.
\newblock {\em Partial differential equations}, volume~19.
\newblock American Mathematical Soc., 2010.

\bibitem{FIDORRA20092142}
M.~Fidorra, A.~Garcia, J.H. Ipsen, S.~Härtel, and L.A. Bagatolli.
\newblock Lipid domains in giant unilamellar vesicles and their correspondence
  with equilibrium thermodynamic phases: A quantitative fluorescence microscopy
  imaging approach.
\newblock {\em Biochimica et Biophysica Acta (BBA) - Biomembranes},
  1788(10):2142 -- 2149, 2009.
\newblock Includes Special Section: Cardiolipin.

\bibitem{Funkhouser_et_al2014}
Chloe~M. Funkhouser, Francisco~J. Solis, and K.~Thornton.
\newblock Dynamics of coarsening in multicomponent lipid vesicles with
  non-uniform mechanical properties.
\newblock {\em The Journal of Chemical Physics}, 140(14):144908, 2014.

\bibitem{gomez2008isogeometric}
H{\'e}ctor G{\'o}mez, Victor~M Calo, Yuri Bazilevs, and Thomas~JR Hughes.
\newblock Isogeometric analysis of the {Cahn--Hilliard} phase-field model.
\newblock {\em {Computer Methods in Applied Mechanics and Engineering}},
  197(49-50):4333--4352, 2008.

\bibitem{Heftberger2015}
P.~Heftberger, B.~Kollmitzer, A.A. Rieder, H.~Amenitsch, and G.~Pabst.
\newblock In situ determination of structure and fluctuations of coexisting
  fluid membrane domains.
\newblock {\em Biophys J.}, 108(4):854--862, 2015.

\bibitem{PMID:20642452}
Janos Juhasz, James~H Davis, and Frances~J Sharom.
\newblock Fluorescent probe partitioning in giant unilamellar vesicles of
  'lipid raft' mixtures.
\newblock {\em The Biochemical journal}, 430(3):415—423, September 2010.

\bibitem{Kahya2003}
N.~Kahya, D.~Scherfeld, K.~Bacia, B.~Poolman, and P.~Schwille.
\newblock Probing lipid mobility of raft-exhibiting model membranes by
  fluorescence correlation spectroscopy.
\newblock {\em J Biol Chem.}, 278(30):28109--15, 2003.

\bibitem{kang2013simple}
You~Jung Kang, Harrison~S Wostein, and Sheereen Majd.
\newblock A simple and versatile method for the formation of arrays of giant
  vesicles with controlled size and composition.
\newblock {\em Advanced Materials}, 25(47):6834--6838, 2013.

\bibitem{KLYMCHENKO201497}
Andrey~S. Klymchenko and Remy Kreder.
\newblock Fluorescent probes for lipid rafts: From model membranes to living
  cells.
\newblock {\em Chemistry \& Biology}, 21(1):97 -- 113, 2014.

\bibitem{C3SM51829A}
Benjamin Kollmitzer, Peter Heftberger, Michael Rappolt, and Georg Pabst.
\newblock Monolayer spontaneous curvature of raft-forming membrane lipids.
\newblock {\em Soft Matter}, 9:10877--10884, 2013.

\bibitem{KUZMIN20051120}
Peter~I. Kuzmin, Sergey~A. Akimov, Yuri~A. Chizmadzhev, Joshua Zimmerberg, and
  Fredric~S. Cohen.
\newblock Line tension and interaction energies of membrane rafts calculated
  from lipid splay and tilt.
\newblock {\em Biophysical Journal}, 88(2):1120--1133, 2005.

\bibitem{Landau_Lifshitz_1958}
L.D. Landau and E.M. Lifshitz.
\newblock {\em Statistical physics}.
\newblock Oxford: Pergamon, 1958.

\bibitem{Levental2020}
I.~Levental, K.R. Levental, and F.A. Heberle.
\newblock Lipid rafts: Controversies resolved, mysteries remain.
\newblock {\em Trends Cell Biol.}, 30(5):341--353, 2020.

\bibitem{Li_et_al2012}
Shuwang Li, John Lowengrub, and Axel Voigt.
\newblock Locomotion, wrinkling, and budding of a multicomponent vesicle in
  viscous fluids.
\newblock {\em Communications in Mathematical Sciences}, 10:645--670, 2012.

\bibitem{Lingwood46}
Daniel Lingwood and Kai Simons.
\newblock Lipid rafts as a membrane-organizing principle.
\newblock {\em Science}, 327(5961):46--50, 2010.

\bibitem{lowengrub2009phase}
John~S Lowengrub, Andreas R{\"a}tz, and Axel Voigt.
\newblock Phase-field modeling of the dynamics of multicomponent vesicles:
  Spinodal decomposition, coarsening, budding, and fission.
\newblock {\em Physical Review E}, 79(3):031926, 2009.

\bibitem{majd2005hydrogel}
Sheereen Majd and Michael Mayer.
\newblock Hydrogel stamping of arrays of supported lipid bilayers with various
  lipid compositions for the screening of drug--membrane and protein--membrane
  interactions.
\newblock {\em Angewandte Chemie}, 117(41):6855--6858, 2005.

\bibitem{ORG09}
Maxim~A Olshanskii, A.~Reusken, and J.~Grande.
\newblock A finite element method for elliptic equations on surfaces.
\newblock {\em SIAM Journal on Numerical Analysis}, 47:3339--3358, 2009.

\bibitem{olshanskii2017trace2}
Maxim~A Olshanskii and Arnold Reusken.
\newblock Trace finite element methods for {PDEs} on surfaces.
\newblock In {\em Geometrically Unfitted Finite Element Methods and
  Applications}, pages 211--258. Springer, 2017.

\bibitem{Palzhanov2021}
Yerbol Palzhanov, Alexander Zhiliakov, Annalisa Quaini, and Maxim Olshanskii.
\newblock A decoupled, stable, and linear fem for a phase-field model of
  variable density two-phase incompressible surface flow.
\newblock {\em Computer Methods in Applied Mechanics and Engineering},
  387:114167, 2021.

\bibitem{park2018reconstitution}
SooHyun Park and Sheereen Majd.
\newblock Reconstitution and functional studies of hamster p-glycoprotein in
  giant liposomes.
\newblock {\em PloS one}, 13(6):e0199279, 2018.

\bibitem{Risselada2008}
H~Jelger Risselada and Siewert~J Marrink.
\newblock The molecular face of lipid rafts in model membranes.
\newblock {\em Proceedings of the National Academy of Sciences},
  105(45):17367--17372, 2008.

\bibitem{SAKUMA20201576}
Yuka Sakuma, Toshihiro Kawakatsu, Takashi Taniguchi, and Masayuki Imai.
\newblock Viscosity landscape of phase-separated lipid membrane estimated from
  fluid velocity field.
\newblock {\em Biophysical Journal}, 118(7):1576--1587, 2020.

\bibitem{Sempkowski2016}
Michelle Sempkowski, Charles Zhu, Monica~Zofia Menzenski, Ioannis~G.
  Kevrekidis, Frank Bruchertseifer, Alfred Morgenstern, and Stavroula Sofou.
\newblock Sticky patches on lipid nanoparticles enable the selective targeting
  and killing of untargetable cancer cells.
\newblock {\em Langmuir}, 32(33):8329--8338, 2016.

\bibitem{B901587F}
Stefan Semrau and Thomas Schmidt.
\newblock Membrane heterogeneity -- from lipid domains to curvature effects.
\newblock {\em Soft Matter}, 5:3174--3186, 2009.

\bibitem{SEZGIN20121777}
Erdinc Sezgin, Ilya Levental, Michal Grzybek, Günter Schwarzmann, Veronika
  Mueller, Alf Honigmann, Vladimir~N. Belov, Christian Eggeling, Ünal Coskun,
  Kai Simons, and Petra Schwille.
\newblock Partitioning, diffusion, and ligand binding of raft lipid analogs in
  model and cellular plasma membranes.
\newblock {\em Biochimica et Biophysica Acta (BBA) - Biomembranes},
  1818(7):1777 -- 1784, 2012.

\bibitem{sohn2010dynamics}
Jin~Sun Sohn, Yu-Hau Tseng, Shuwang Li, Axel Voigt, and John~S Lowengrub.
\newblock Dynamics of multicomponent vesicles in a viscous fluid.
\newblock {\em Journal of Computational Physics}, 229(1):119--144, 2010.

\bibitem{Stanich2013}
Cynthia~A Stanich, Aurelia~R Honerkamp-Smith, Gregory~Garbes Putzel,
  Christopher~S Warth, Andrea~K Lamprecht, Pritam Mandal, Elizabeth Mann,
  Thien-An~D Hua, and Sarah~L Keller.
\newblock Coarsening dynamics of domains in lipid membranes.
\newblock {\em Biophysical journal}, 105(2):444--454, 2013.

\bibitem{Trementozzi2019}
Andrea~N. Trementozzi, Zachary~I. Imam, Morgan Mendicino, Carl~C. Hayden, and
  Jeanne~C. Stachowiak.
\newblock Liposome-mediated chemotherapeutic delivery is synergistically
  enhanced by ternary lipid compositions and cationic lipids.
\newblock {\em Langmuir}, 35(38):12532--12542, 2019.

\bibitem{van1984ostwald}
RA~Van~Santen.
\newblock The ostwald step rule.
\newblock {\em The Journal of Physical Chemistry}, 88(24):5768--5769, 1984.

\bibitem{veatch2003separation}
Sarah~L Veatch and Sarah~L Keller.
\newblock Separation of liquid phases in giant vesicles of ternary mixtures of
  phospholipids and cholesterol.
\newblock {\em Biophysical journal}, 85(5):3074--3083, 2003.

\bibitem{Veatch17650}
Sarah~L. Veatch, Olivier Soubias, Sarah~L. Keller, and Klaus Gawrisch.
\newblock Critical fluctuations in domain-forming lipid mixtures.
\newblock {\em Proceedings of the National Academy of Sciences},
  104(45):17650--17655, 2007.

\bibitem{Balint2017}
\v{S}. B\'alint and M.L. Dustin.
\newblock Localizing order to boost signaling.
\newblock {\em eLife}, 6:e25375, 2017.

\bibitem{Wang2008}
Xiaoqiang Wang and Qiang Du.
\newblock Modelling and simulations of multi-component lipid membranes and open
  membranes via diffuse interface approaches.
\newblock {\em Journal of Mathematical Biology}, 56(3):347--371, Mar 2008.

\bibitem{Wesolowska2009}
O.~Wesolowska, K.~Michalak, J~Maniewska, and A.~B. Hendrich.
\newblock Giant unilamellar vesicles - a perfect tool to visualize phase
  separation and lipid rafts in model systems.
\newblock {\em Acta Biochimica Polonica}, 56:33--39, 2009.

\bibitem{Yushutin_IJNMBE2019}
Vladimir Yushutin, Annalisa Quaini, Sheereen Majd, and Maxim Olshanskii.
\newblock A computational study of lateral phase separation in biological
  membranes.
\newblock {\em International journal for numerical methods in biomedical
  engineering}, 35(3):e3181, 2019.

\bibitem{zhiliakov2021experimental}
Alexander Zhiliakov, Yifei Wang, Annalisa Quaini, Maxim Olshanskii, and
  Sheereen Majd.
\newblock Experimental validation of a phase-field model to predict coarsening
  dynamics of lipid domains in multicomponent membranes.
\newblock {\em Biochimica et Biophysica Acta (BBA)-Biomembranes},
  1863(1):183446, 2021.

\bibitem{PhysRevE.60.3564}
Jingzhi Zhu, Long-Qing Chen, Jie Shen, and Veena Tikare.
\newblock Coarsening kinetics from a variable-mobility {Cahn-Hilliard}
  equation: Application of a semi-implicit {Fourier} spectral method.
\newblock {\em Phys. Rev. E}, 60:3564--3572, Oct 1999.

\end{thebibliography}

\pagebreak

\begin{center}
\Large{Supplementary material to the article titled: ``Lipid domain coarsening and fluidity in multicomponent lipid vesicles: A continuum based model and its experimental validation''}
\end{center}

\vskip 1cm 

To estimate the surface density of each phase, we first calculated the molar-weighted average values for the molecular weight ($MW$) and the molecular surface area ($A$) for the corresponding phase, and calculated the surface density by:
\begin{equation*}
\rho = \frac{MV}{A}.
\end{equation*}
The average $MW$ for each phase was calculated using its lipid composition, 
as summarized in Table 2, and the molecular weight of individual lipids (785.593 g/mol for DOPC, 
733.562 g/mol for DPPC and 386.65 g/mol for chol). 
Similarly, the average value of $A$ for each phase was found based on its lipid composition 
and cross-sectional area of its lipid components. The cross-sectional area of lipids were estimated 
using an approach detailed in our previous study [46]. Table S1 summarizes the calculated values.

\begin{table}[ht]
\renewcommand\thetable{S1}
\centering
\begin{tabular}{ |c|c|c|c|c|c|c|}
 \hline
Membrane composition & \multicolumn{2}{| c |}{Average $MV$}  & \multicolumn{2}{| c |}{
Average $A$} &  \multicolumn{2}{| c |}{$\rho = MV/A$} \\
 \hline
DOPC:DPPC:Chol (Temp) & $l_d$ & $l_o$ & $l_d$ & $l_o$ & $l_d$ & $l_o$ \\
\hline
1:2:25\% (15\textdegree{}C) & 742.231 & 635.732 & 63.156 & 43.346 & $1.951e-23$ & $2.435e-23$\\
\hline
1:2:25\% (17.5\textdegree{}C) & 741.190 & 640.242 & 63.271 & 44.618 & $1.945e-23$ & $2.382e-23$\\
\hline
1:1:15\% (22.5\textdegree{}C) & 742.057 & 649.609 & 63.300 & 46.366 & $1.946e-23$ & $2.326e-23$ \\
\hline
1:1:15\% (25\textdegree{}C) & 739.587 &651.689 & 62.479 & 47.603 & $1.965e-23$ & $2.273e-23$ \\
\hline
\end{tabular}
\caption{The average molecular weight ($MW$) and molecular surface area ($A$) of lipids in each phase, 
and the surface density of the corresponding phase ($\rho$) in the examined membrane compositions 
at different temperatures. Molecular weight ($MW$) is presented in Kg/mol, $A$ in $\r{A}$, and $\rho$ in g/$\r{A}^2$.}\label{tab:thermo}
\end{table}%

\end{document}